\documentclass [] {aa}
\usepackage{graphicx}
\usepackage{natbib}
\bibpunct{(}{)}{;}{a}{}{,}

\input psfig.sty

\def \grss {GRS 1915+105~}
\def \grs {GRS 1915+105}

\def \sax {{\it Beppo}SAX}

\begin{document}
%\thesaurus{08.14.1; 08.16.7 PSR~B1509$-$58; 13.25.5)}
\title{The complex behaviour of the microquasar GRS~1915+105 in the $\rho$
class observed with \sax. II: Time-resolved spectral analysis}
\author{T.~Mineo\inst{1}
\and E.~Massaro\inst{2}
\and A.~D'Ai\inst{3}
\and F.~Massa\inst{4,5}
\and M.~Feroci\inst{6}
\and G.~Ventura\inst{4}
\and P.~Casella\inst{7}
\and C.~Ferrigno\inst{8}
\and T.~Belloni\inst{9}
}
\institute{INAF, Istituto di Astrofisica Spaziale e Fisica
cosmica di Palermo, via U. La Malfa 153, I-90146 Palermo, Italy
\and Dipartimento di Fisica, Universit\`a La Sapienza, Piazzale A. Moro 2,
I-00185 Roma, Italy
\and Dipartimento di Fisica, Universit\`a di Palermo, Via Archirafi 36,
I-90123 Palermo, Italy
\and Stazione Astronomica di Vallinfreda, via del Tramonto, Vallinfreda (RM), Italy
\and INFN-Sezione di Roma1 (retired), Roma, Italy
\and INAF, Istituto di Astrofisica Spaziale e Fisica
cosmica di Roma, via del Fosso del Cavaliere 100, I-00113 Roma, Italy
\and School of Physics and Astronomy, University of Southampton, Southampton, SO17 1BJ, 
United Kingdom
\and ADSC, Data Centre for Astrophysics, Chemin d'Ecogia 16, CH-1290 Versoix
Switzerland
\and INAF, Osservatorio Astronomico di Brera, via E. Bianchi 46, I-23807 Merate,
Italy
}
\offprints{Teresa Mineo: \\
mineo@iasf-palermo.inaf.it}
\date{Received ....; accepted ....}
\markboth{T.~Mineo et al.: The complex behaviour of GRS~1915+105 in the $\rho$ class.
II: Spectral analysis}
{T.~Mineo et al.: The complex behaviour of GRS~1915+105 in the $\rho$ class.
II: Spectral analysis}

\abstract{{\it Beppo}SAX observed GRS 1915+105 on October 2000 with a long pointing
lasting about ten days. During this observation, the source was mainly in the
$\rho$ class characterized by bursts with a recurrence time of between 40 and 100 s.
} { 
We identify five segments in the burst structure and accumulate the average spectra 
of these segments during each satellite orbit.
We present a detailed spectral analysis aimed at determining
variations that occur during the burst and understanding the physical process
that produces them.
} {  
We compare MECS, HPGSPC, and PDS spectra  with several models. 
Under the assumption that a single model is able to fit all spectra,
we find that the combination of a multi-temperature
black-body disk and a hybrid corona is able to give a consistent physical
explanation of the source behaviour.
} {   
Our measured variations in $KT_{el}$, $\tau$, $KT_{in}$, and $R_{in}$
appear to be  either correlated or anti-correlated
with the count rate in the energy range 1.6-10 keV.  
The strongest variations are detected along the burst segments:
almost all parameters exhibit significant variations  
in the segments that have the highest fluxes ($pulse$)
with the exception of $R_{in}$, which varies continuously and reaches a
maximum just before the peak.
The flux of the multi-temperature disk strongly increases
in the $pulse$ and simultaneously the corona contribution is
significantly reduced.  
} { 
The disk luminosity increases in the $pulse$ and the
$R_{in}-T_{in}$ correlation can be most successfully interpreted 
in term of the slim disk model.
In addition, the reduction in the corona luminosity at the bursts
might represent the condensation of the corona onto the disk.
} {
}  
\authorrunning{T. Mineo et al.}
\titlerunning{The complex behaviour of GRS~1915+105 in the $\rho$ class.}

\maketitle

%%%%%%%%%%%%%%%%%%%%%%%%%%%%%%%%%%%%%%%%%%%%%%%%%%%%%%%%%%%%%%%%%%%%%%%%%%%%%%%%

\section{Introduction}

The microquasar \grss \citep{Fender2004} was discovered in 1992 in the hard X-ray band
\citep{Castro-Tirado1992}, and later identified with a binary system of an orbital
period of about 33 days \citep{Greiner2001} containing a black hole whose estimated
mass is 14.0$\pm$4.4 $M_{\odot}$ \citep{Harlaftis2004}.
Its distance, evaluated by  \citet{Mirabel1994} and later revised by
\citet{Fender1999}, is about 12.5 kpc, and
a source inclination of 66\degr -- 70\degr is inferred
from the apparent superluminal motion observed in the radio jet
\citep{Mirabel1994,Fender1999}.

The source has been extensively observed in the X-ray band 
\citep[][]{Fender2004,Rodriguez2008,Ueda2010,Vierdayanti2010}.
The X-ray emission is characterized by strong variability
with alternating  bright bursts and quiet phases.
\citet{Belloni2000} classified the X-ray temporal and spectral behaviour of \grss
into 12 classes, and  \citet{Klein-Wolt2002} and \citet{Hannikainen2005}
identified additional sets of variability patterns.
As in the case of many black hole candidates (BHC), the spectrum can generally 
be fitted with at least two components:
a first one, dominating at soft X-ray energies ($\leq$10 keV), interpreted in terms
of thermal emission, and a second one extending  
up to several hundreds keV that can be modeled with a power law.
The thermal component, which has a higher temperature  than in other
BHCs, has been interpreted as emission from an accretion
disk surrounding the black hole. According to the most accreditated
hypothesis, the disk is a thin standard one extending up to a few tens of 
kilometers from the central source \citep{Fender2004}.
However, the presence of a slim disk \citep{Abramowicz1988} has also been
advocated to describe some of the spectra observed from this source
\citep{Ueda2009,Vierdayanti2010,Yadav1999}.
The phenomenological power-law  model is used to describe a Comptonization
component from an optically thick corona
above the disk with an electron population containing either thermal
or non-thermal particles \citep{Zdziarski2001}
and where energy can be also gained from the gas bulk motion  
\citep{Titarchuk1997}.
In addition to the main spectral components, lower intensity features
have  occasionally been detected: a Compton reflection from the disk
\citep{Zdziarski2001,Ueda2009}, an iron line present only in
a few observations \citep{Feroci1999, Martocchia2002, Martocchia2006, Ueda2010}.

During the bursts, the main spectral parameters of the thermal component
exhibit significant variations, which have been interpreted in term of the emptying
and refilling of the inner portion of the accretion disk \citep{Belloni1997}.
In agreement with the phenomenology observed in other black holes,
the spectral evolution of \grss in all classes can been described as transitions
between three separate states, as suggested by \citet{Belloni2000} from the
analysis of RXTE observations: a state {\bf A} with a small time variability
where a strong black-body like component with a temperature $>$1 keV dominates
the spectrum, a state {\bf B} with a disk temperature that is higher that in state
{\bf A} and a red-noise time variability on a timescale $>$1 s, and
state {\bf C} where the spectrum is dominated by a power law with a spectral
index between 1.8 and 2.6 and white and red noise time variability on a timescale $>$1 s.

The \sax~ satellite observed \grss for about ten days in October 2000.
For almost the entire duration of the pointing, the source was in the
$\rho$ class according to the classification of \citet{Belloni2000},
characterized by quasi-regular bursts with a slow rise and a sharp decrease.
We described the time evolution of the source `heartbeat' activity in this
period in  \citet[][hereafter Paper I]{Massaro2010}, where, on
the basis of Fourier and wavelet analysis, we defined the {\it regular} and
{\it irregular} modes of this class.
The former mode is characterized by periodograms with one or two
prominent peaks and a relatively small variance in the dominant timescale
in the wavelet scalograms.
In the irregular mode, periodograms exhibit several peaks and wavelet spectra
have highly fluctuating timescales.
Moreover, pulses of regular series have a number of peaks ({\it multiplicity})
that is generally no greater than two, whereas in the bursts of irregular series it
increases even to more than four.
The two modes were rather clearly separated in time: the irregular mode
belongs mainly to the central part of observation (170-375 ks from the
beginning), whereas the regular mode was observed in the first
and the last portions.

% ========= Table 1 =====================

\begin{table*}[h]
\caption{Log of the {\it Beppo}SAX runs used for the analysis.
The columns list the observation codes, the name of the runs,
the time in seconds from the starting time of the
first observation (20 October 2000 at UT = 21$^h$ 26$^m$ 55$^s$),
the exposure time, the rate observed in the energy range
used in the analysis, and the number of run bursts.}
\label{table1}
\begin{center}
\begin{tabular}{cclccccc}
\hline\hline
Obs. code & Run & Tstart & Exposure & \multicolumn{3}{c}{Rate } & N burst \\
& & (s) & (s) & \multicolumn{3}{c}{ (ct s$^{-1}$)}&\\
& & & & MECS & PDS & HP &\\
\hline
21226001 & A2b & 8994.0 & 2191.0 & 203.1 & 39.6 & 80.2 & 43 \\
& A3 & 14727.0 & 2509.5 & 203.9 & 39.2 & 82.6 & 53 \\
& A4 & 20648.0 & 3152.0 & 198.0 & 36.5 & 80.1 & 66 \\
& A5 & 26511.0 & 2809.5 & 193.4 & 35.1 & 79.1 & 57 \\
& A6 & 31928.5 & 3108.0 & 191.8 & 35.4 & 77.2 & 64 \\
& A7 & 37662.0 & 3156.5 & 199.2 & 38.4 & 79.8 & 63 \\
& A8b & 43402.0 & 2507.0 & 197.2 & 38.4 & 78.7 & 62 \\
& A9 & 49296.0 & 2952.5 & 206.1 & 41.9 & 82.5 & 56 \\
212260011 & B2b & 94999.0 & 2197.0 & 207.6 & 40.6 & -- & 45 \\
& B3 & 100740.0 & 2638.5 & 199.6 & 38.0 & -- & 55\\
& B4 & 107594.0 & 2008.0 & 207.6 & 40.5 & -- & 38\\
& B5 & 112199.0 & 3093.0 & 212.2 & 42.5 & -- & 61 \\
& B6 & 117933.0 & 3146.5 & 208.3 & 40.7 & -- & 65\\
& B7 & 123666.5 & 3122.0 & 199.3 & 38.6 & -- & 62\\
& B8 & 129400.0 & 3057.5 & 203.3 & 39.2 & -- & 61\\
& B9b & 136123.5 & 2130.0 & 205.8 & 41.0 & -- & 42\\
& B10 & 141260.0 & 2685.0 & 210.1 & 41.6 & -- & 52\\
212260012 & C1 & 181003.0 & 2198.0 & 203.4 & 36.5 & 83.8 & 34\\
& C2 & 186736.5 & 2679.0 & 206.1 & 41.1 & 84.0 & 49\\
& C3 & 192470.0 & 3103.5 & 206.6 & 39.2 & 83.3 & 61\\
& C4 & 198203.5 & 3121.0 & 207.3 & 40.8 & 83.6 & 61\\
& C5 & 203937.0 & 3086.5 & 210.9 & 43.0 & 84.6 & 59\\
& C6 & 209816.5 & 2942.0 & 211.6 & 40.2 & 86.0 & 59 \\
& C7 & 215404.0 & 3067.5 & 212.5 & 39.7 & 88.1 & 60 \\
212260013 & D5b & 267005.0 & 2219.0 & 215.3 & 40.9 & 86.3 & 43\\
& D6 & 272748.0 & 2675.0 & 211.3 & 39.7 & 85.5 & 54 \\
& D9 & 289938.0 & 3117.5 & 228.7 & 39.7 & 96.5 & 50\\
& D10 & 295671.0 & 3129.5 & 227.8 & 39.0 & 96.8 & 45\\
& D11 & 301404.5 & 2987.5 & 232.2 & 39.4 & 98.0 & 39\\
& D12 & 307138.5 & 3158.0 & 225.8 & 40.2 & 98.4 & 40\\
& D13 & 313230.0 & 2730.5 & 216.5 & 39.1 & 91.8 & 40\\
212260014 & E4 & 358737.5 & 2666.5 & 221.3 & 38.9 & 91.5 & 39\\
& E5 & 364470.5 & 3063.0 & 212.9 & 37.5 & 88.7 & 47\\
& E6 & 370204.0 & 3025.5 & 204.7 & 39.6 & 85.3 & 42\\
& E7 & 375937.0 & 3112.5 & 201.9 & 40.8 & 83.1 & 49\\
& E8 & 381670.5 & 3091.0 & 212.3 & 42.7 & 84.8 & 59\\
& E10 & 393136.5 & 3062.0 & 215.8 & 44.2 & 84.9 & 55\\
& E12 & 405336.0 & 2361.5 & 235.1 & 51.2 & 94.7 & 36\\
212260015 & F1 & 439021.0 & 2194.0 & 243.0 & 52.5 & 97.9 & 31\\
& F2 & 444735.5 & 2644.5 & 231.3 & 48.9 & 92.3 & 40\\
& F3 & 450468.0 & 3020.0 & 231.9 & 49.5 & 93.6 & 45 \\
& F4 & 456201.0 & 3102.0 & 241.8 & 51.6 & 96.3 & 47\\
& F6 & 467752.0 & 2979.5 & 237.8 & 50.6 & 95.1 & 49\\
& F7 & 473400.5 & 3099.0 & 234.3 & 49.4 & 93.0 & 51\\
& F8 & 479133.5 & 2934.5 & 233.6 & 49.2 & 92.5 & 46\\
& F9 & 485200.0 & 2744.5 & 229.6 & 48.7 & 91.5 & 44\\
& F17 & 530768.0 & 2578.0 & 244.3 & 51.9 & 97.8 & 36 \\
212260016 & G3 & 547929.0 & 3070.5 & 243.0 & 52.1 & -- & 46\\
& G4 & 553723.5 & 3026.0 & 234.1 & 48.4 & -- & 50\\
& G5 & 559395.0 & 3072.5 & 235.6 & 48.9 & -- & 53\\
& G6b & 566000.5 & 2153.0 & 244.1 & 53.0 & -- & 33\\
& G7 & 571145.0 & 2781.5 & 230.0 & 48.4 & -- & 42\\
% & G16b & 629170.5 & 2070.0 & 269.4 & 59.2 & -- & 21 \\
\hline
\end{tabular}
\end{center}

\end{table*}

In this paper, we present a time-resolved spectral analysis of a large fraction of
the data series of Paper I with the principal aim of obtaining more information on
the spectral changes associated with the bursts.
We included, in addition, results from a short simultaneous RXTE observation.
A spectral analysis of the $\rho$ mode was presented by \citet{Taam1997} and \citet{Vilhu1998}
using data from RXTE. Nevertheless, a complete study of the burst processes, which is typical
of the $\rho$ class, is not yet available
(while this paper was under the referee process \citet{Neilsen2011} published their results on the 
analysis of $\rho$ mode spectra observed with RXTE).

In our analysis, we  applied to the data set several spectral models that have
been found to closely fit the energy spectral distributions of \grss in various
variability classes; we identified those able to provide a satisfactory general description
of the spectra, and then studied the variations in their main properties.
Moreover, we present the results of a much shorter simultaneous RXTE observation.

Data and selection procedures are illustrated in Sects. 2 and 3, the spectral analysis is
presented in Sect. 4, and the time evolution of the spectral parameters is given in Sect. 5.
The analysis of RXTE data is described in Sect. 6 and the physical discussion of the
results is provided in Sect. 7.

\section{Observation and data reduction}

{\it Beppo}SAX observed the \grss in 2000 from October 20 (MJD 51837.894) to
October 29.
As in Paper~I, we used the data obtained with the Medium Energy Concentrator
Spectrometer (MECS), operating in the 1.3--10~keV energy band \citep{Boella1997},
and the Phoswich Detector System (PDS) operating in the
15--300~keV energy band \citep{Frontera1997}.
Furthermore, we considered the data obtained with the High Pressure Gas
Scintillation Proportional Counter \citep[HPGSPC;][]{Manzo1997},
which worked in the nominal energy band 4--120 keV, and covers the gap between
the two other instruments.

Data retrieved from {\it Beppo}SAX archive at the ASI Science Data Center
are organized into observations containing several satellite orbits
with a period of $\sim$96 m.
In Paper~I, we divided the observations into runs, each corresponding to
a continuous observing period, and named the runs with capital letters and
sequential numbers. In this paper, we present the spectral
analysis performed over runs having exposure times greater than 2000 s
in order to have, in all instruments, a number of counts sufficient to confine
the spectral parameters.
Table~\ref{table1} lists these 52 runs, together with the code of the observation
from where they were extracted, the time in seconds from the starting time
of the first observation (20 October 2000 at UT = 21$^h$ 26$^m$ 55$^s$),
the exposure, the rate of each instrument in the
energy range considered for the spectral analysis (1.6--10 keV for the MECS,
8--30 keV for the HPGSPC and 15--150 keV for the PDS),
and the number of bursts detected in each series.

The extraction and reduction procedures for the MECS and PDS data were already
described in Paper I; for HPGSPC, we used the latest distributed version of
SAXDAS (v2.3.1) and applied the standard selection criteria.
The tool used to extract HPGSPC spectra ({\sc hpproducts} v3.1.1) produces some
failure in the gain correction of data from observations 212260011 and 212260016,
which were consequently excluded from the analysis.

The MECS background, estimated from high Galactic latitude `blank' fields
accumulated in the same region of the detectors, is neglected,
as it has a count rate of a level 2.6$\times$10$^{-2}$ c s$^{-1}$, which is much lower
than the typical source flux, even in the faintest states.
Both the PDS and HPGSPC background were obtained from simultaneous measurements thanks to
the standard operational rocking mode.

Our spectral analysis was performed with XSPEC v.11 using the response matrices,
that are available from the {\it Beppo}SAX archive\footnote{http://www.asdc.asi.it/bepposax}.
We restricted the spectral analysis of HPGSPC data to the energy range 8--30 keV
where the response matrix is more accurate.
The MECS-PDS intercalibration factor was fixed to 0.85, which is its average value
in the range 0.83-0.89 for point sources of high count rate,
and the MECS-HPGSPC factor was limited to vary within the range 0.9-1.1
to take into account the discrepancy in the flux measured by the two instruments.
We found that HPGSPC measured a flux $\sim$7\% higher
than the MECS.
 
The errors in the values of the best-fit parameters reported throughout the paper and in the
tables correspond to a 1 $\sigma$ confidence level for a single parameter, unless
explicitly stated.

\section{Burst ``anatomy'' and count rate variations}

\begin{figure}[t]
\vspace{0.2 cm}
\centerline{
\psfig{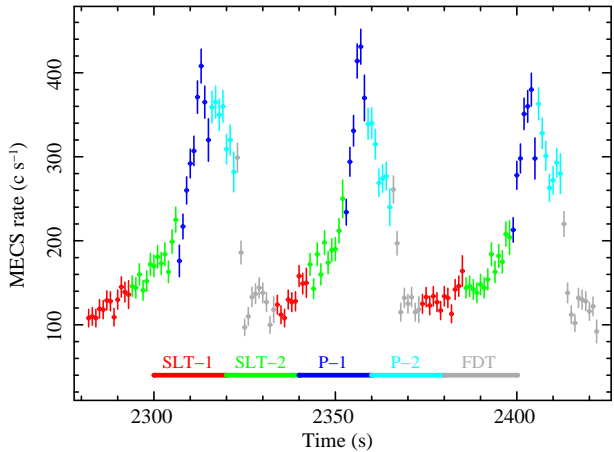}
}
\caption{A snap shot of the light curve in the entire MECS energy range
of the series A3. Red and green data points indicate the {\it SLT-1} and{\it SLT-2}
intervals,
blue and cyan points are for {\it P-1} and {\it P-2}, and grey points indicate the ending part
of the burst ({\it FDT}).}
\label{figura1}
\end{figure}

In Paper I, we distinguished three main  phase intervals or segments of the burst
profile and investigated the time evolution of the count rates for each of them.
The most prominent feature is the {\it pulse}, which  generally consist of several 
short peaks.
Prior to the {\it pulse}, there is the {\it slow leading trail} (hereafter {\it SLT}), 
which is also called the `shoulder' in the literature, during which the count 
rate increases much slower than in the {\it pulse}.
There is no clear separation between these two components, and we took it at the
time where the count rate reached the level of the FWHM of the {\it pulse},
which was evaluated by means of a Gaussian best-fit.
Between the end of the {\it pulse} and the beginning of the {\it SLT},
the count rate reaches its lowest level that can be considered as a {\it baseline}
upon which bursts are superposed.
We defined a time interval of three seconds where it is possible to estimate
this {\it baseline level} (hereafter {\it BL} corresponding to
the first three seconds of the red segment of the light curve shown in Fig.~\ref{figura1}) and
found that {\it BL} remained remarkably stable
in each individual time series (see Paper I for more details).
In the spectral analysis, we essentially followed this division and introduced a finer
partition.

The  analysis of the {\it BL} spectra gave parameters in agreement,
within the errors, with those of the {\it SLT} ones, and owing to the
lower quality statistics, we did not consider {\it BL} as a separate segment.
The signal in the entire {\it SLT} is strong enough to divide this component into
two parts of equal length, hereafter called {\it SLT-1} and {\it SLT-2}, respectively,
which are plotted using data points of different colors in Fig.~\ref{figura1}, where a short
segment of the series A3 is reported as an example.
The same colors are also used in the other figures to permit an easier identification of
the various time segments.
We also divided the {\it pulse} into two equal time intervals, namely
{\it P-1} and {\it P-2}, which are indicated in Fig. 1 and in subsequent figures
by blue and cyan data points, respectively.
An even finer segmentation to enable us to analyze individual peaks was not possible
because the peak shape and duration is highly variable from burst to burst.
Finally, the ending part of the pulse to the initial minimum of {\it SLT} was taken as a
fifth segment, marked in Fig. 1 with light grey points, and labeled {\it FDT} 
corresponding to {\it Final Decaying Trail}.
The starting and ending times of these parts in each burst were evaluated by means of
the same procedure described in Paper I.
For each data series of MECS, HPGSPC (whenever available), and PDS, we accumulated
five spectra, each of them corresponding to the integral of a selected part of the
burst.

%Figura 1

\begin{figure}[ht]
\vspace{0.2 cm}
\centerline{\psfig{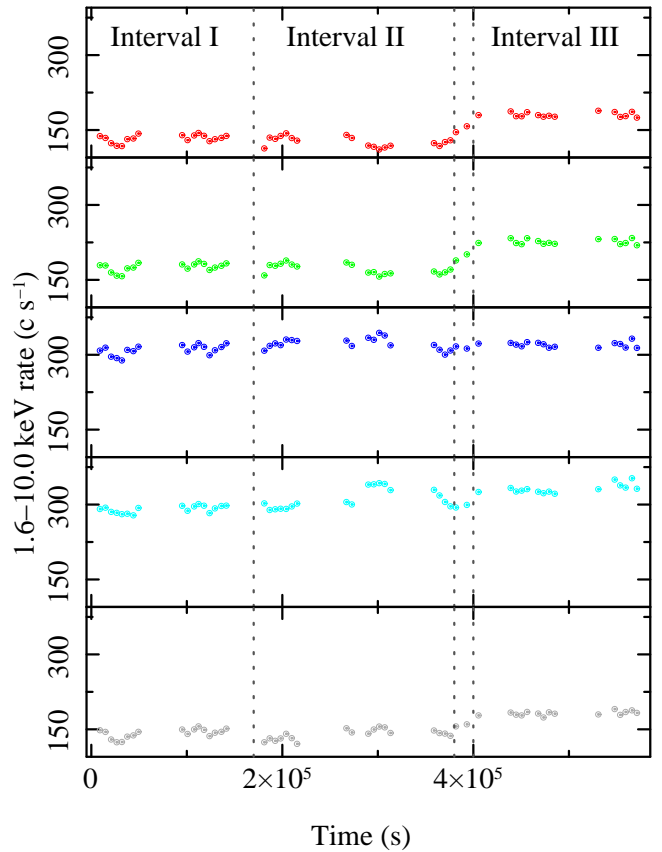}}
\caption{Time history of MECS (1.6--10 keV) count rates for the
five segments identified along the bursts.
Top to bottom:
{\it SLT-1} (red), {\it SLT-2} (green), {\it P-1} (blue), {\it P-2}
(cyan), and {\it FDT} (grey).
Vertical dashed lines mark the main intervals of the observation.
}
\label{mrate}
\end{figure}

\begin{figure}[ht]
\vspace{0.2 cm}
\centerline{\psfig{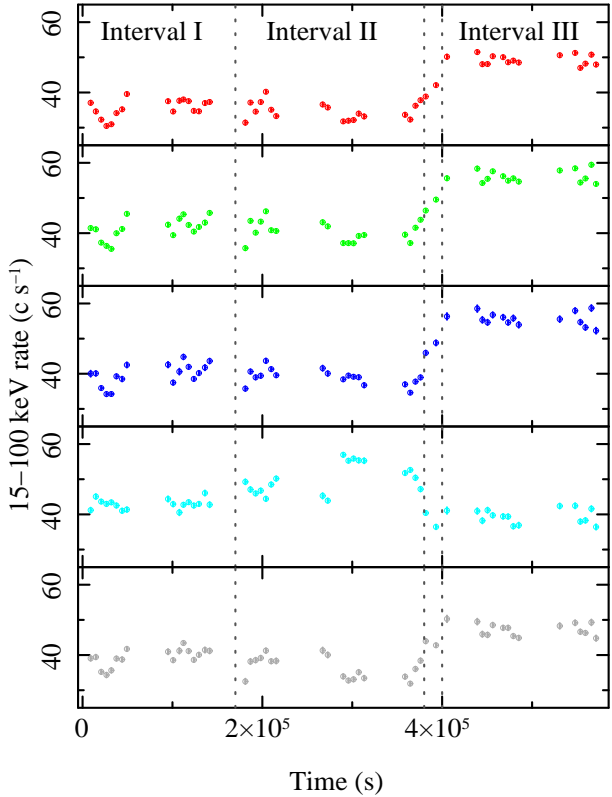}}
\caption{Time history of PDS (15--100 keV) count rates for the  
five  segments identified along the bursts.
Top to bottom:
{\it SLT-1} (red), {\it SLT-2} (green), {\it P-1} (blue), {\it P-2}
(cyan), and {\it FDT} (grey).
Vertical dashed lines mark the main intervals of the observation.
}
\label{pdsrate}
\end{figure}

The variations in the count rates of the various components in the course of this
long pointing were analysed in detail in Paper I.
However, to take into account the finer segmentation of the bursts' profile introduced
during the spectral analysis, we again tracked  the behaviour of each part to 
highlight  any possible differences from  the results presented in Paper I.
We considered the same time segmentation of the whole dataset in three intervals
introduced in Paper I: interval I from the start time to 1.7$\times$10${^5}$ s,
interval II from this time to 3.8$\times$10${^5}$ s, and the last interval III is from
4.0$\times$10${^5}$ s to 6.0$\times$10${^5}$ s.
In particular, the second interval is characterized by the presence of the {\it irregular}
mode detected in the timing analysis (Paper I), and the last interval corresponds to the final
high state after the rapid increase in count rate detected between 3.8$\times$10${^5}$ s and
 4.0$\times$10${^5}$ s.
Figures ~\ref{mrate} and ~\ref{pdsrate} show the average rate for the five portions
of the bursts for each series in the 1.6--10 keV (MECS) and 15--100 keV (PDS) energy
ranges, respectively.

%Figura 2

As already discussed in Paper I, the evolution of the count rates is not the same for
the different parts of the bursts.
The behaviours of {\it SLT-1}, {\it SLT-2}, and {\it FDT} are similar in the MECS and PDS 
data sets, and their count rates undergo 
a significant increase between 380 ks and 400 ks with nearly stable levels before
and after this transition, as was also found for the {\it BL} level (see the lower panel
of Fig.15 in Paper I).

%Figura 3

In the MECS data, the count rates of {\it P-1} and {\it P-2} remain quite stable
during the entire length of the observation, without any indication of a step
increase.
We note that the relative {\it pulse} contribution in the MECS range is higher in
the first 400 ks of observation (before the step increase) as shown in Fig. 15
of Paper I, where we present the difference of the rate with respect to the BL level. 
In the PDS data, the rate variation among the five segments is lower
than in the MECS (the largest difference is 1.7 compared to 2.5 in the MECS),
and the presence of the peak contribution is only evident in
{\it P-2}, for  which there is  a strong increase in the rate during the first 400 ks
with a bump in the central part of the pointing where the source was in  an
irregular mode.
The hardening of {\it P-2} during the irregular mode was already reported in Paper I,
where we noted that PDS peaks were clearly apparent only in the final part of
the {\it pulse}, while in the regular series they were not so high.
In summary, there is a strong correlation between the MECS and the
PDS rate, which weakens at the $pulse$, where the MECS rate is almost constant.

The spectral variations within the five segments correspond to cyclical changes in the  
state defined by \citet{Belloni2000}: the source is mainly in state {\bf C} in
the {\it SLT}, moves to {\bf A} in {\it P-1}, then to {\bf B} in {\it P-2},
and, finally,  returns to {\bf C} in the {\it FDT} section.

\section{Spectral analysis}

As simple spectral models had been shown not to give acceptable fits over an
energy range from a fraction of a keV to about one hundred keV,
we considered composite models.
The majority of the models that we studied had indeed been previously
 used to investigate this source,
although for different brightness states and variability classes.
These models include the emission from a standard optically thick accretion disk 
to reproduce the spectrum up to about 15 keV \citep[{\sc diskbb} in XSPEC][]
{Mitsuda1984}, and, at higher energies, they consider other spectral functions  
that generally have a significant contribution also at lower energies.
We tried several models (listed in Appendix A) that
interpret the expected spectral shape either with a combination  of phenomenological
functions or with laws based on theoretical %self-consistent
descriptions of a corona-accretion disk  system.  

%figura 4
\begin{figure}[h]
%\vspace{0.5 cm}
\centerline{
\vbox{
\psfig{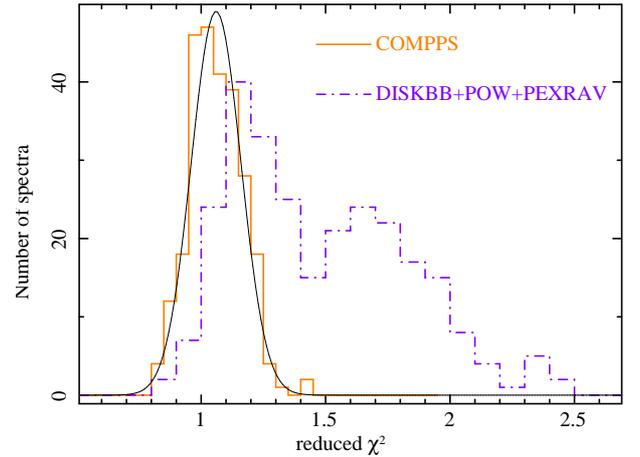}
}}
\caption{Frequency histograms of the reduced $\chi^2$ for all spectra of the ``good
model'' (continuous orange line) and of model $iv$ described in Appendix A
(dashed line) compared with the expected distribution evaluated by a MonteCarlo
simulation (black line).
}
\label{chi_bb}
\end{figure}

We required that the same spectral model was able to fit all the 52 spectra
of each segment by  changing only the values of the parameters.
To test the consistency of the model with the data, we compared the frequency
distribution of the reduced $\chi^2$ with the expected parent distribution.
This was evaluated from 10000 simulated distributions each generated by randomizing
a set of 52 values from the $\chi^2$ functions corresponding to the degree of
freedom of each observed spectrum.
We found that a model closely fitting all spectra has an expected distribution
of the reduced $\chi^2$ given by a Gaussian-like function centered on 1.00$\pm$0.02
that has a standard deviation of $\sigma$=0.10$\pm$0.01.
As a result of our fits, we found that the reduced $\chi^2$
distribution has the expected width (0.10$\pm$0.01), but the mean (1.05$\pm$0.01)
is shifted even if compatible with the expected one to within two standard deviations.
This shift is mainly caused by  the highest quality data statistics 
(such as those of the $pulse$) 
and may be produced by residual systematics originating either from local detector 
features at the maximum level of 5\% (such as the 4.78 keV residual in the MECS spectra) 
or from the uncertainty in the intercalibration factors ($\sim$3\%) fixed for the PDS to 0.85
(see Sect.~2)\footnote{http://http://www.asdc.asi.it/bepposax/software/cookbook}.
This effect cannot be properly treated within
XSPEC, where systematics are introduced as a single factor
over the entire fitting range. To take this effect into account, 
we compared our $\chi^2$ distribution with a Gaussian
centered on 1.06 and the compatibility of the resulting
distributions with the expected one was confirmed by means
of a Kolmologov-Smirnov test.

%figura 5
\begin{figure*}[!th]
\centerline{
\vbox{
\hbox{
\hspace{-1.0cm}
\psfig{figure=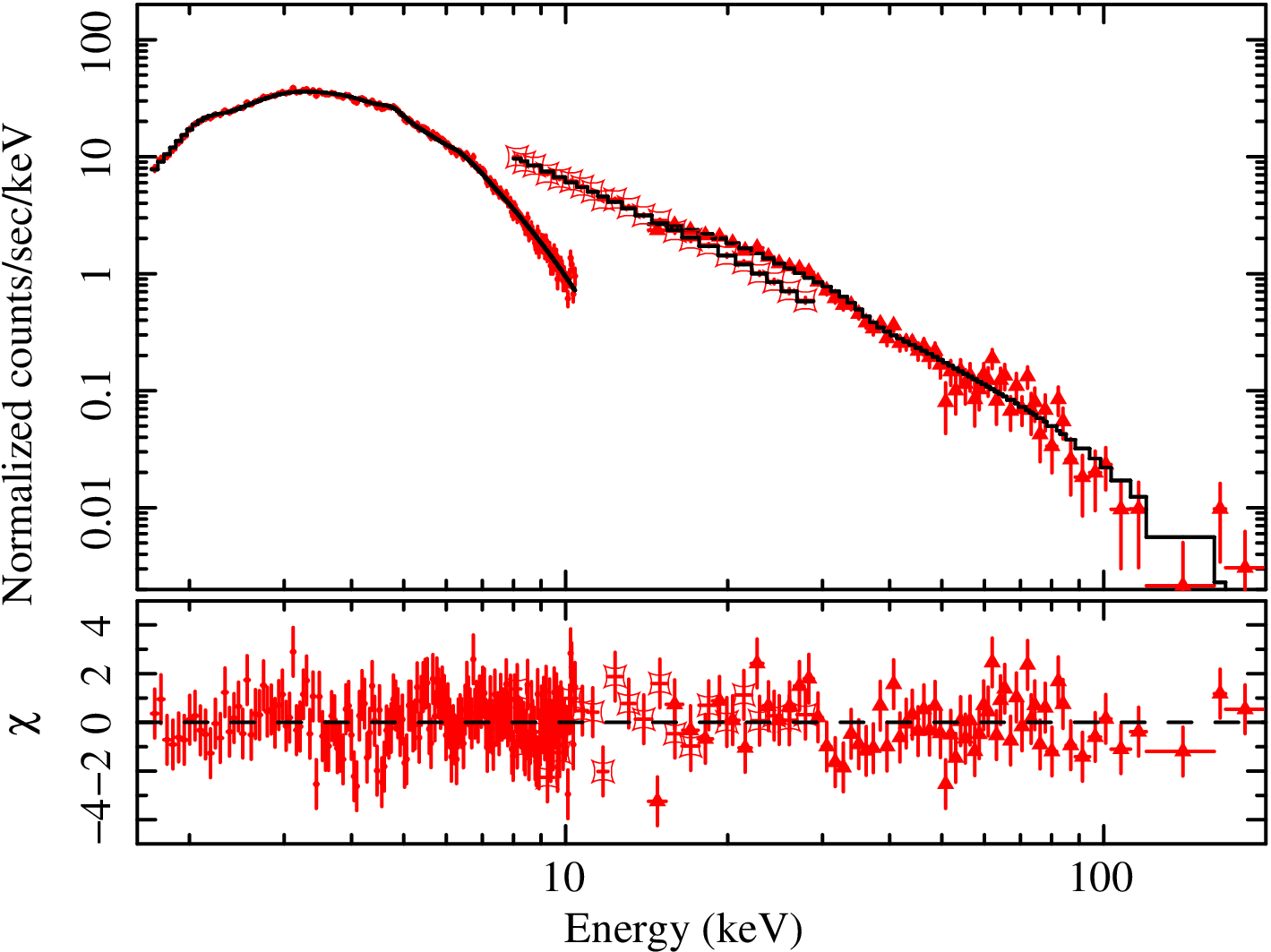,width=6.0cm,angle=-90,clip=}
\hspace{2.0cm}
\psfig{figure=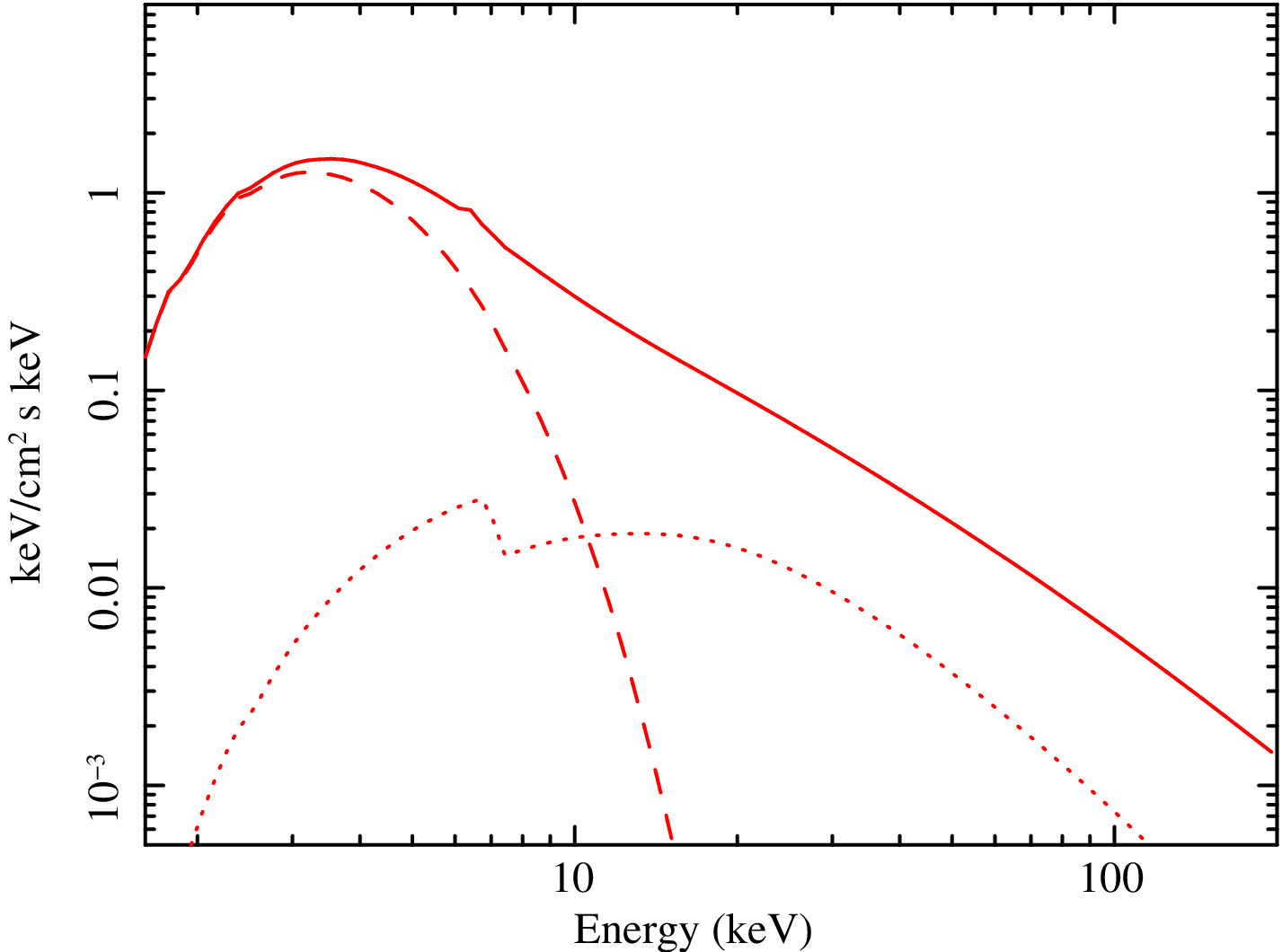,width=6.0cm,angle=-90,clip=}
}
\vspace{0.3 cm}
\hbox{
\hspace{-1.0cm}
\psfig{figure=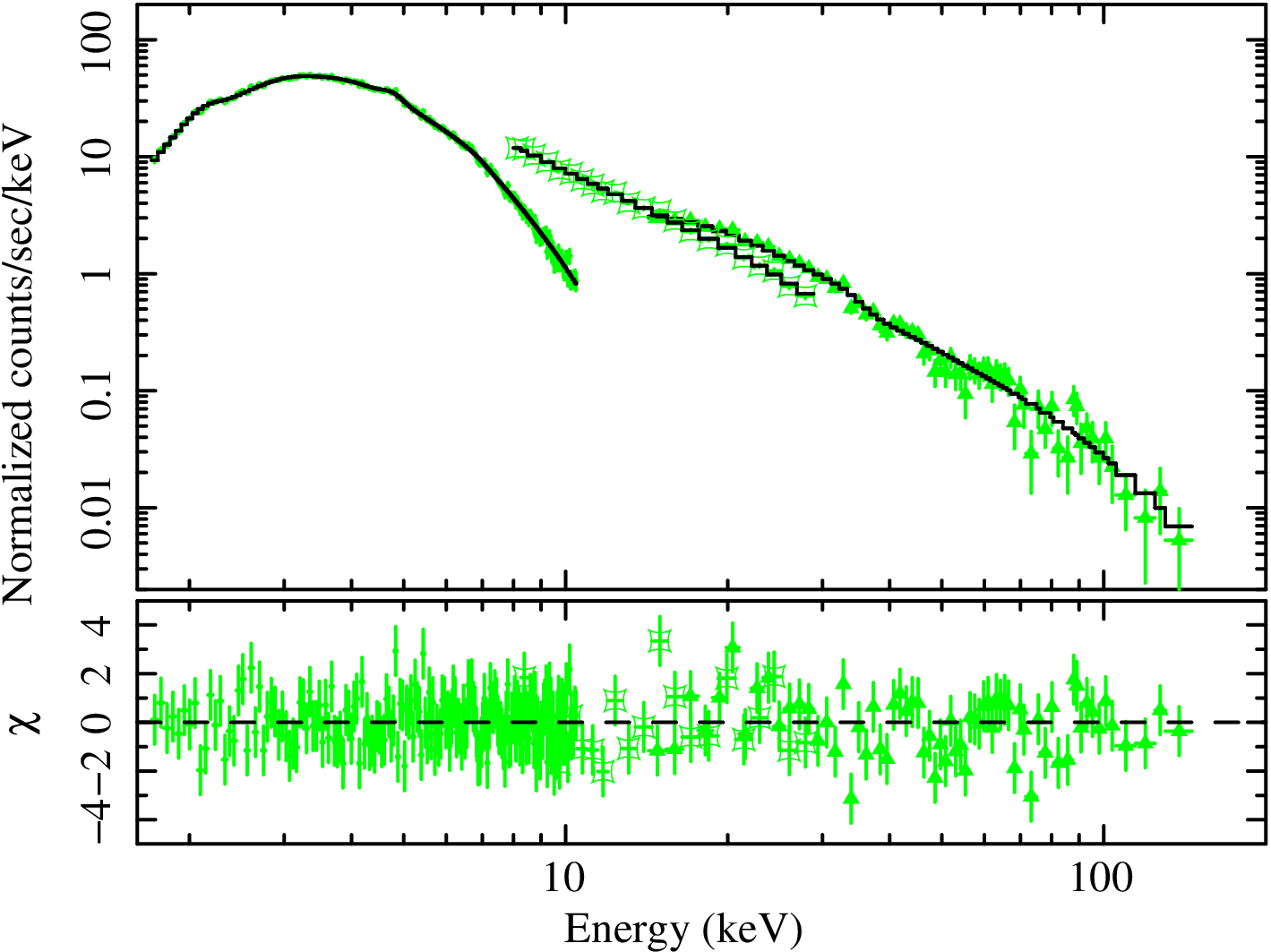,width=6.0cm,angle=-90,clip=}
\hspace{2.0cm}
\psfig{figure=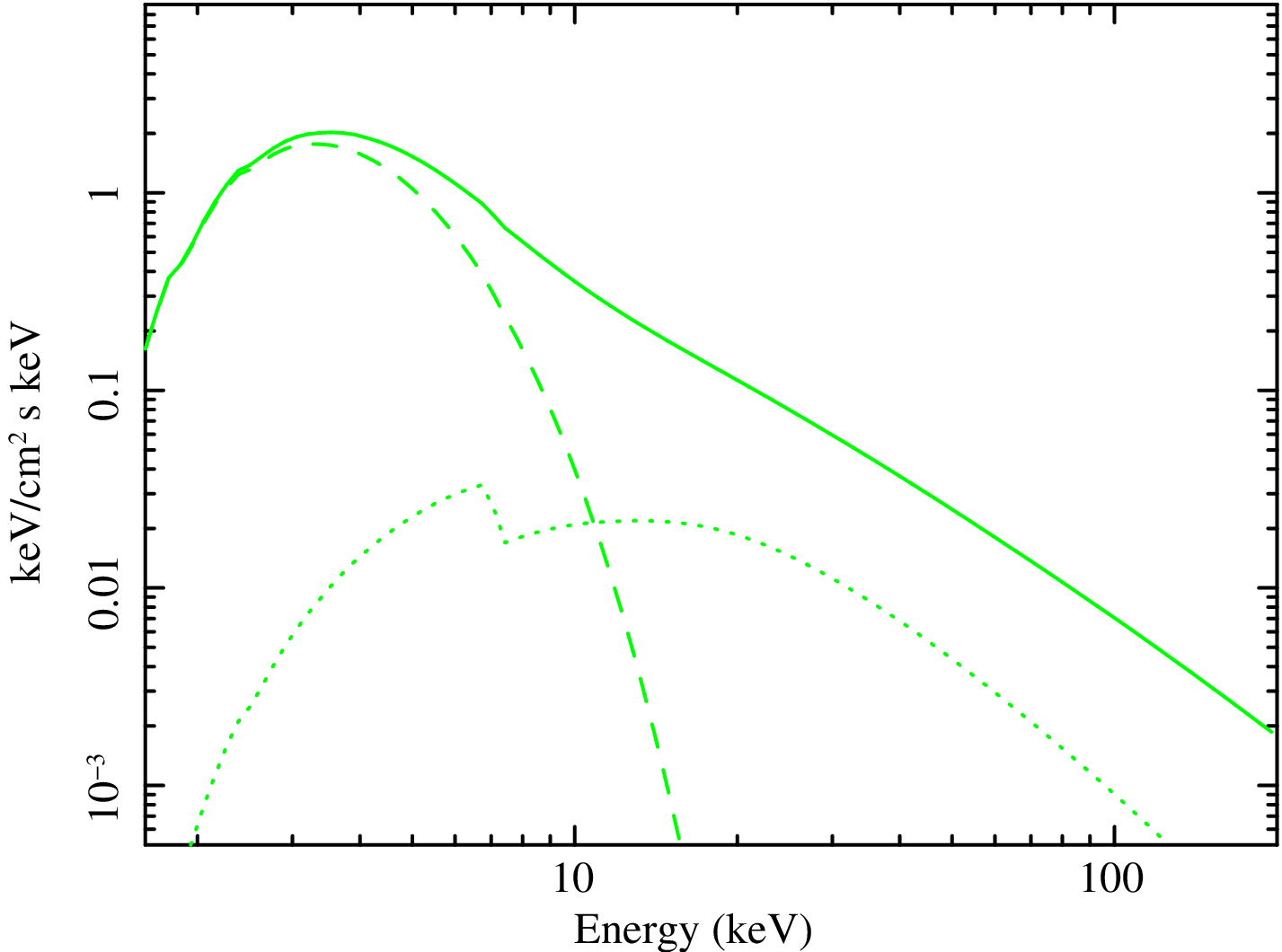,width=6.0cm,angle=-90,clip=}
}
\vspace{0.3 cm}
\hbox{
\hspace{-1.0cm}
\psfig{figure=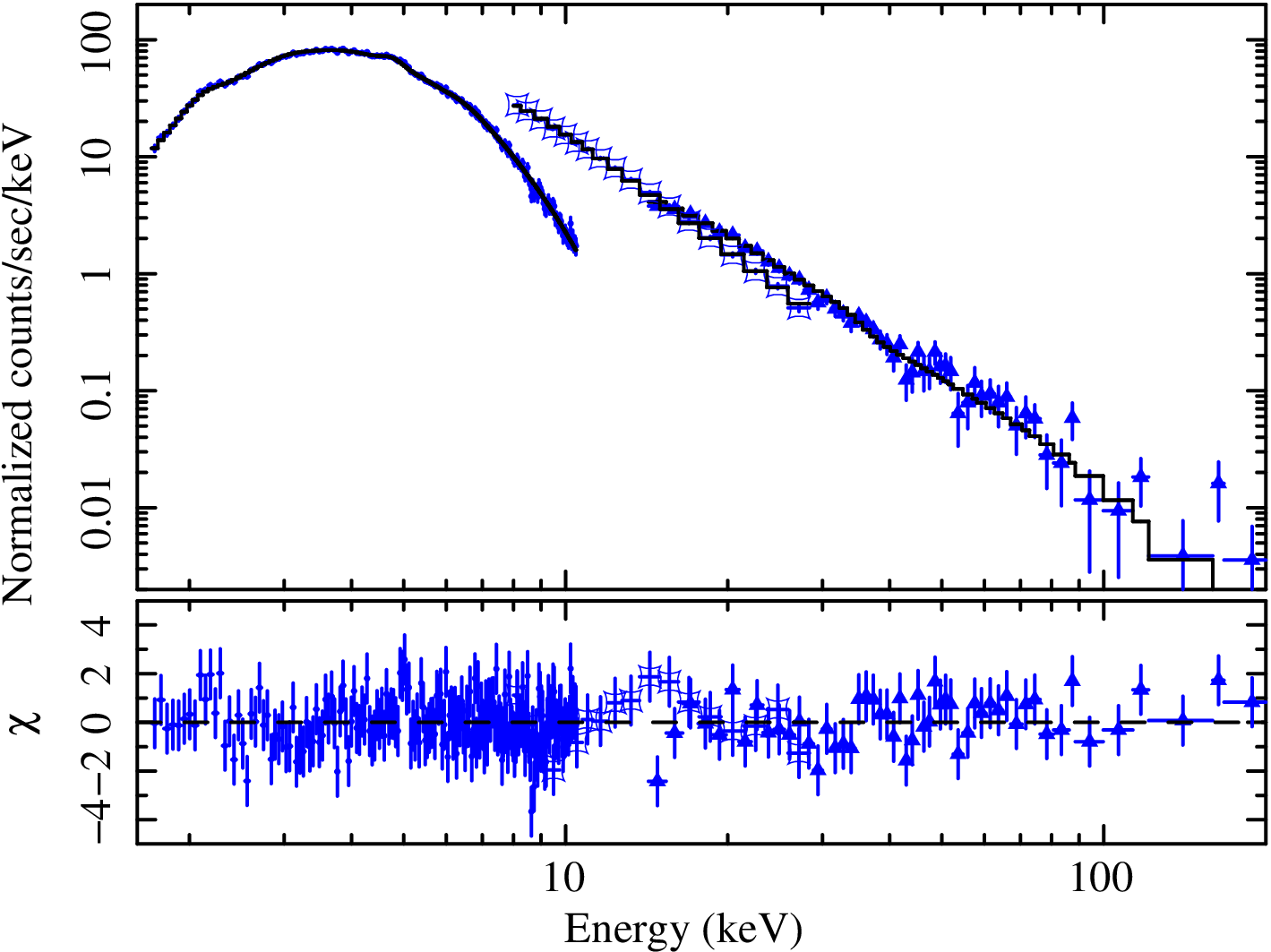,width=6.0cm,angle=-90,clip=}
\hspace{2.0cm}
\psfig{figure=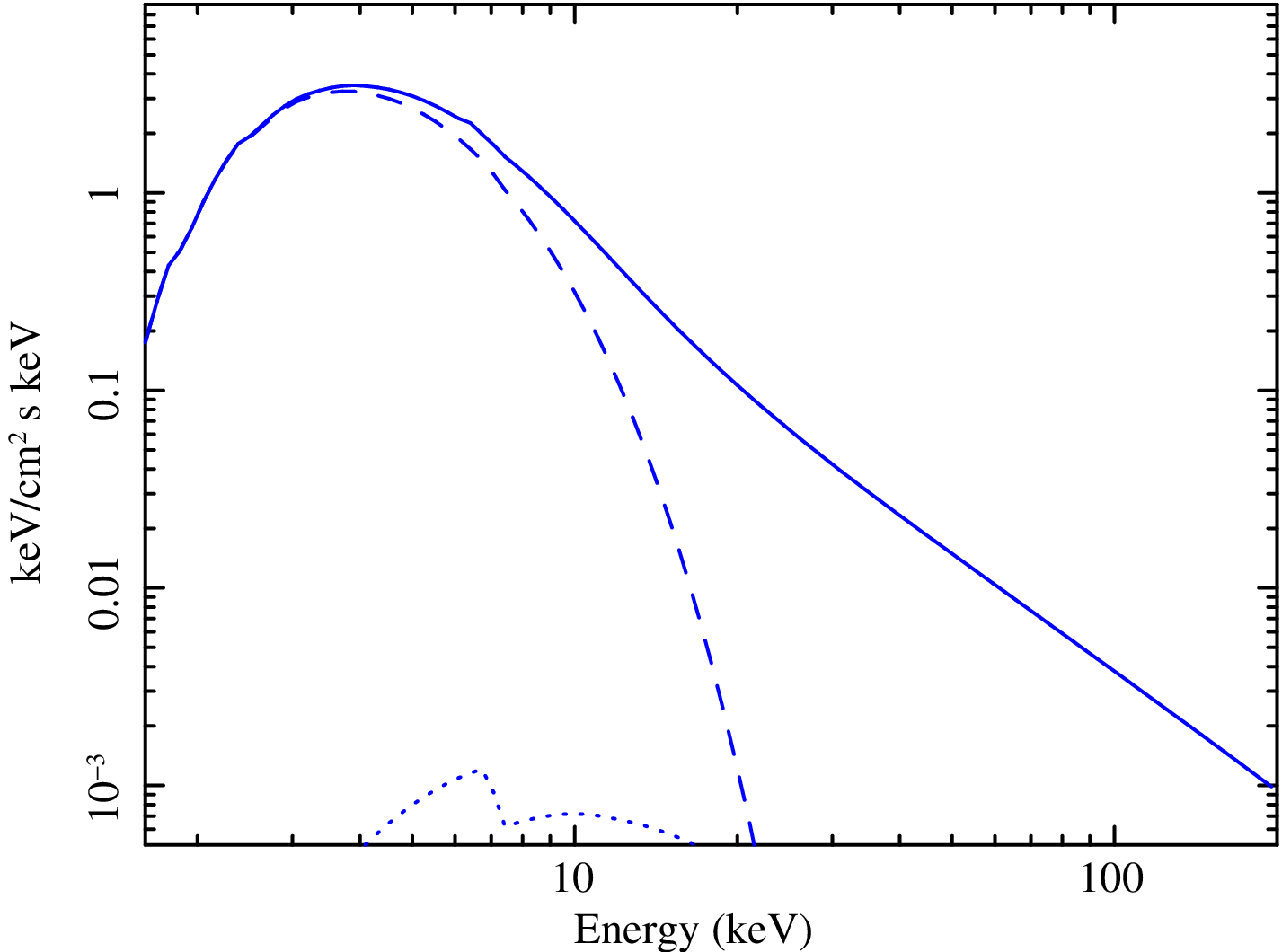,width=6.0cm,angle=-90,clip=}
}
\vspace{0.3 cm}
\hbox{
\hspace{-1.0cm}
\psfig{figure=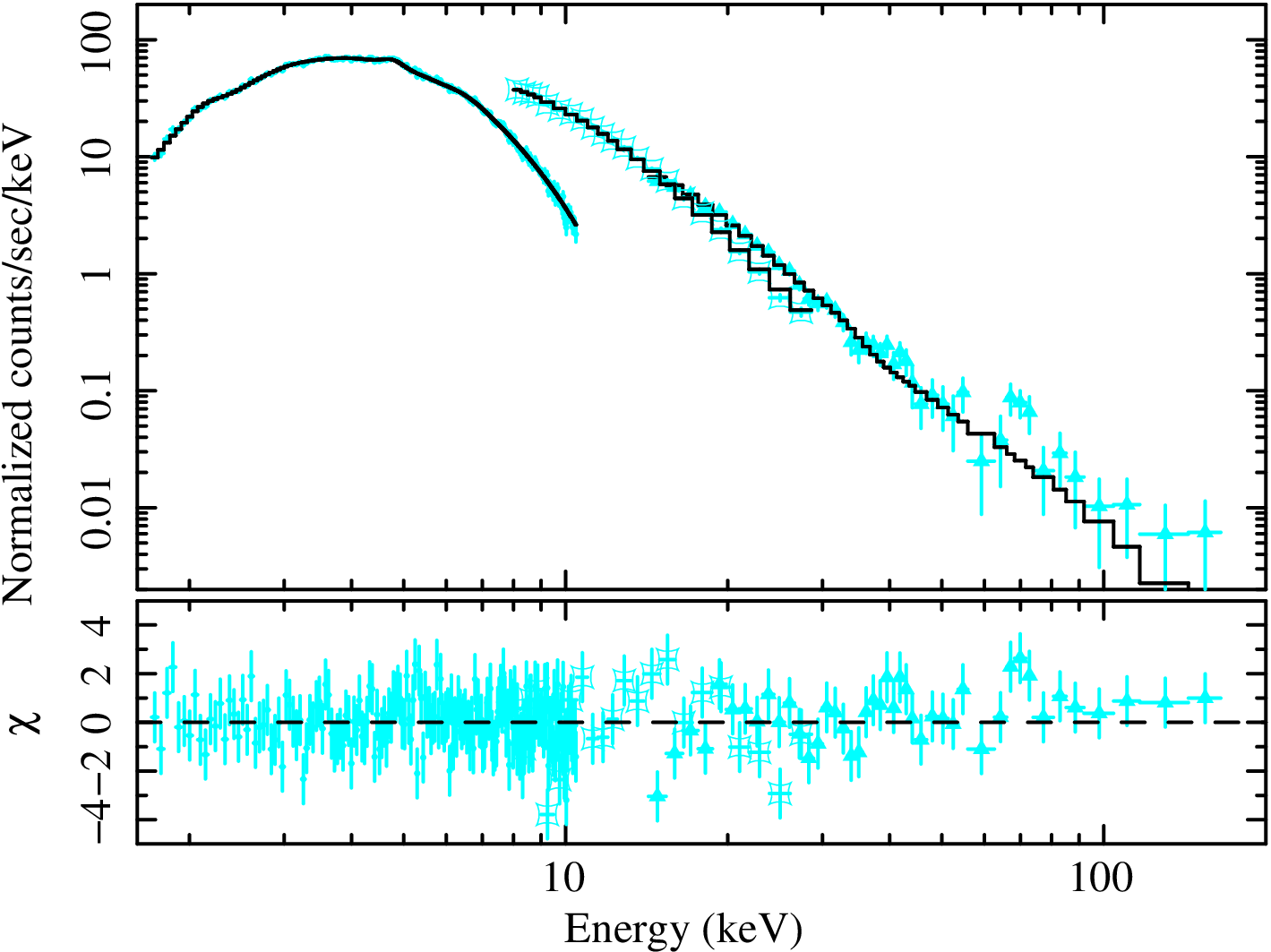,width=6.0cm,angle=-90,clip=}
\hspace{2.0cm}
\psfig{figure=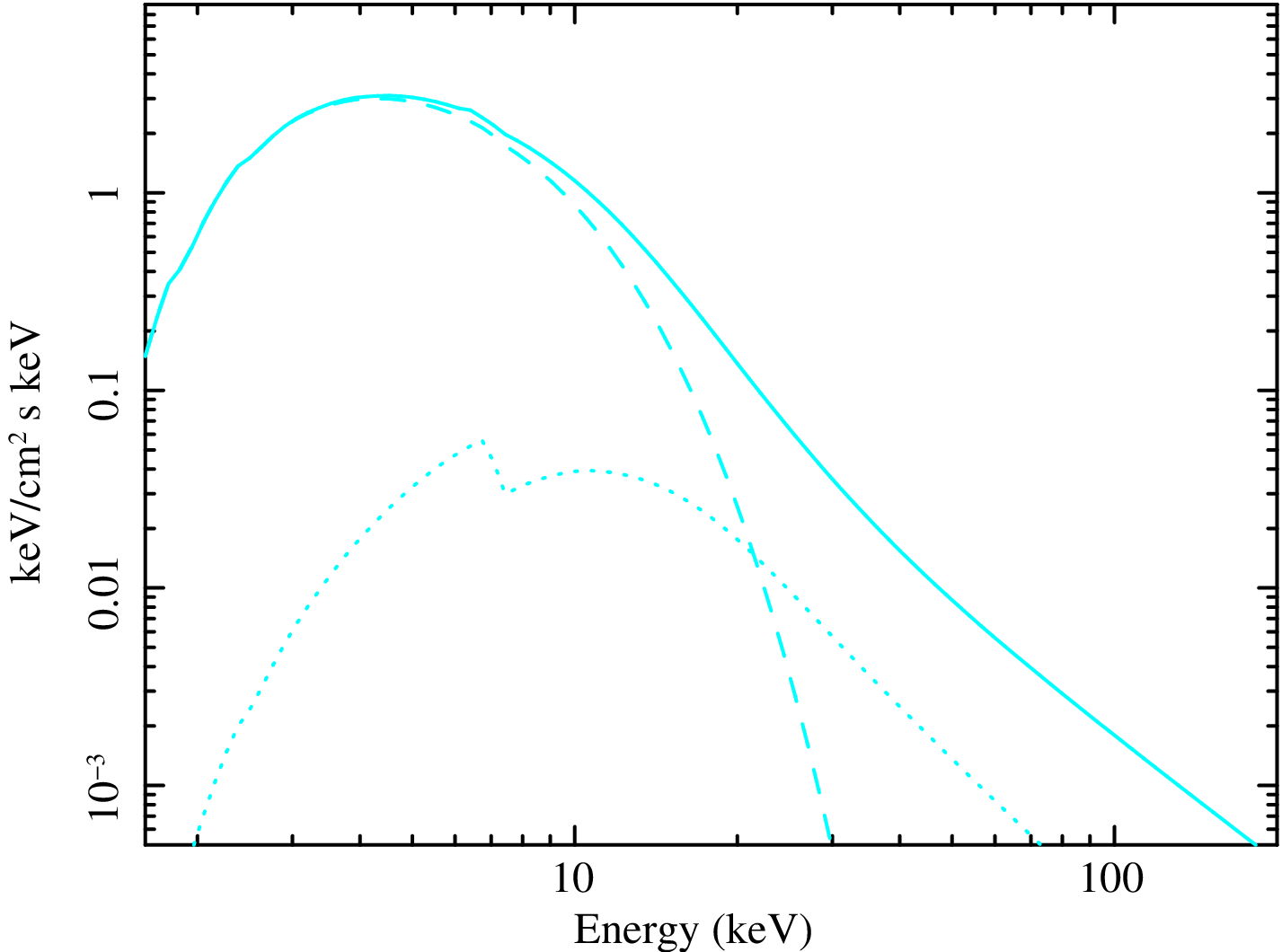,width=6.0cm,angle=-90,clip=}
}
\vspace{0.3 cm}
\hbox{
\hspace{-1.0cm}
\psfig{figure=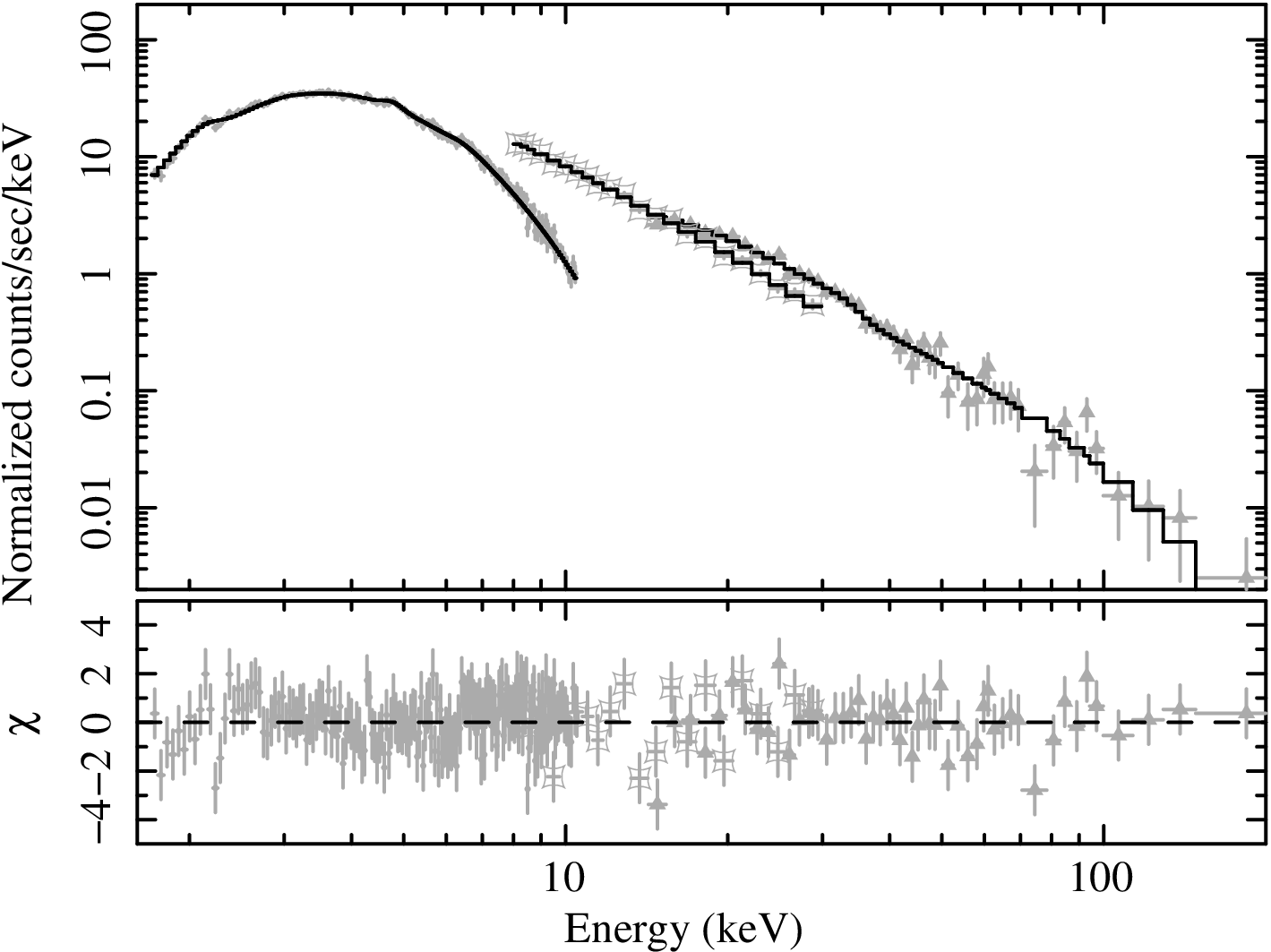,width=6.0cm,angle=-90,clip=}
\hspace{2.0cm}
\psfig{figure=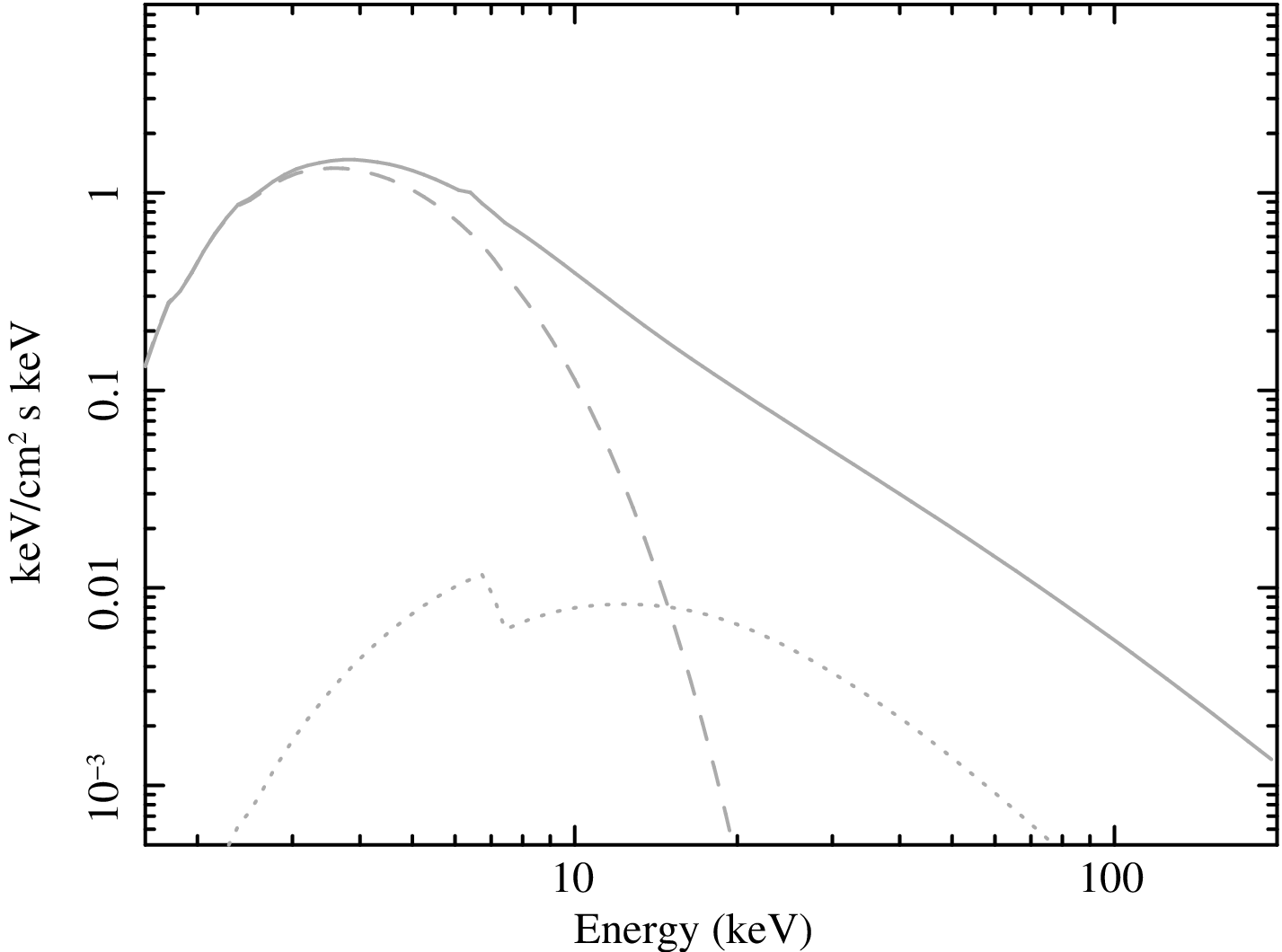,width=6.0cm,angle=-90,clip=}
}
}}
\caption{{\it Left panels:} Data and fitting model plus residuals
for spectra in run E7.
{\it Right panels:} Best-fit models (continuous line) with the
relative spectral components: disk (dashed line) 
and Compton reflection (dotted line).}
\label{spt_bb}
\end{figure*}

We found that the ``good model'' that most closely describes the source behaviour
during the burst and provides a consistent physical interpretation is
the standard disk plus  a corona that contains an hybrid
population of electrons including both thermal and non-thermal
distributions  \citep{Poutanen1996}.
The code relative to this model implemented  in XSPEC ({\sc compps}) 
includes, together with emission from the multi-temperature disk and corona, 
the Compton reflection from
gas whose temperature and ionization status can be set as fitting parameters.
It provides acceptable best-fits to the spectra of all data sets and the
reduced $\chi^2$ distribution is compatible with the expected one within only
one standard deviation as shown in  Fig.~\ref{chi_bb}, where the {\sc compps}
$\chi^2$ distribution is compared with the one relative to the
model $iv$ (top panel) described  in  Appendix A.
The model $v$ in Appendix A ({\sc diskbb+pow+bbody}) also provides fully
acceptable best-fits, without any prominent features in the residuals, 
for the spectra of all data sets.
However, as explained in Appendix A, no simple physical interpretation can be
given for the black-body component, and therefore it was not considered in the 
subsequent analysis.

In the {\sc compps} fits, the electron distribution is assumed to be thermal below a  Lorenz
factor $\gamma_{min}$ above which energies are distributed with a power law spectrum
of index $\Gamma$ up to a Lorenz factor $\gamma_{max}$ fixed at 10$^3$ \citep{Zdziarski2005}.
The values of $ \gamma_{min}$ and $\Gamma$,  which were free to vary during first
sets of fitting, were found to maintain almost constant values
($\gamma_{min}$: mean 1.3, rms 0.7; $\Gamma$: mean 4.6, rms 1.6)
in all the observations and in the various segments and were then
fixed to their average values. The $\gamma_{min}$ closely agrees
with the results obtained by \citet{Zdziarski2005} for the analysis of
RXTE$+$OSSE observations, while, $\Gamma$ is fixed to a value slightly
larger or marginally compatible with those from previous measurements
\citep{Zdziarski2005,Ueda2010}.

The disk, surrounded by a spherical corona (param $geom$=0 in XSPEC),
extends to between 10 and 1000 Schwarzschild radii with an inclination of 
$\theta$=70\degr.
Compton reflection from a  non-rotating disk (param {\it Betor10}=-10 in XSPEC)
with a gas temperature of 10$^6$ K
and a low  level of ionization (param $\xi$=0) is also taken into account.

A reduction in the free parameters led to a broader $\chi^2$ distribution
that was still compatible with the expected one (see Fig.~\ref{chi_bb}).
The free-fitting parameters were thus: the absorbing column density ($N_{\rm H}$),
the temperature ($kT_{el}$), and the optical depth ($\tau$) of the corona,
then the temperature ($kT_{in}$) and the apparent inner radius ($R_{in}$) 
of the accretion disk, and the solid angle of the cold reflecting matter 
to the illuminating source ($\Omega$) expressed in units of 2$\pi$. 
The values of $R_{in}$  are computed from  the normalization of the model, 
$K$,  using the relation $K=(R_{in}[{\rm km}]/D[10\,{\rm kpc}])^2 \times cos(\theta)$, 
where $D$ is the distance (12.5 kpc). They are practically equivalent
(to within a few percent) to the ones computed
using the formula given by \citet{Kubota2004b} that, assuming isotropic radiation,
takes into account the conservation of the photon number in the Comptonization 
process.

The spectral fittings and residuals for the five segments of the series E7
are shown in the left panels of Fig.~\ref{spt_bb}; their relative models
plus components are plotted in the right panels of the same figure.

\subsection{The iron line}
The detection of a fluorescence K$\alpha$ iron line from \grss has been reported
on several occasions \citep{Martocchia2002, Zdziarski2005, Martocchia2006,Ueda2010} and,
more recently, by \citet{Titarchuk2009}.
In particular, \citet{Martocchia2002} from the detection of a strong
line found evidence of relativistic effects in a black hole environment.

We searched for the presence of the K$\alpha$ iron line by including in the adopted
model a Gaussian profile, whose central energy was allowed to vary across the 
range between 6.3 keV and 7.1 keV and  whose  width was fixed to 0.1 keV.
No significant line was observed in any of the analyzed spectra with
a maximum detection of 2.5 $\sigma$ only in a few spectra
and  a 3$\sigma$ upper limit of $\sim$50 eV to the EW.

\subsection{The low energy absorption}
The low energy absorption of \grss is complex and variable
\citep{Belloni2000,Lee2002,Martocchia2006,Yadav2006,Ueda2009,Ueda2010,vanOers2010}.
The values of the equivalent column density $N_{\rm H}$ obtained from the
analysis of X-ray data ranges between 2$\times10^{22}$  and 16$\times10^{22}$ cm$^{-2}$
\citep{Greiner1996,Taam1997,Trudolyubov2001,Martocchia2002,Lee2002,Yadav2006,
vanOers2010} and are generally higher than the
values  derived from the 21 cm hydrogen line emission
along the line of sight \citep[4.7$\pm$0.2$\times10^{22}$ cm$^{-2}$;][]
{Chaty1996} \citep[$\geq1.8\times10^{22}$ cm$^{-2}$;][]{Dickey1990}.
These results indicate that there is  a composite local absorber containing
 materials in grains whose characteristics have been derived with 
Chandra observations  \citep{Lee2002, Ueda2009}. 

Since our data are relative to an instrument of moderate energy resolution,
and a low energy boundary at $\sim$2 keV, we adopted the XSPEC {\sc wabs} function
assuming standard solar abundance from \citet{Anders1989}.
No significant variation in the absorbing column was discerned during the ten
days of observations. However, we found a rather weak increase
in the column density at the peak: the $N_{\rm H}$ varied between
4.3$\times10^{22}$ cm$^{-2}$ in the $SLT$ and 5.4$\times10^{22}$ cm$^{-2}$
in the {\it pulse} (see Table~\ref{table6}).
Fixing the $N_{\rm H}$ to an intermediate value resulted only in
a small worsening of the $\chi^2$ distribution without affecting the values of the
other parameters. In particular, fixing to 5.0$\times10^{22}$ cm$^{-2}$
the $N_{\rm H}$ of the {\it P-2} spectra, differences smaller than 10\% were measured
for the disk parameters and values that were compatible to within 2$\sigma$ for the corona ones.

Following \citet{Ueda2010}, we evaluated the contribution of the halo produced  
by dust scattering to the MECS spectra below 3 keV. We found that it produces about 10\%  
of the emission assuming the fractional intensity 
derived for GX 13$+$1 \citep{Smith2002}. This could account for the apparent variability
of $N_{\rm H}$ on a long timescale, but the scattered flux is not expected 
to vary during the period 40-100 s coherently with the burst,
where the time delay respect to the source emission is about 100 hours \citep{Ueda2010}.

We also fitted our spectra with a more complex absorption model ({\sc tbvarabs} in XSPEC)
by fixing the abundances to the values obtained by \citet{Lee2002} and  \citet{Ueda2009}, and
in both cases  a broader $\chi^2$ distribution centered at $\sim$1.2 was obtained.
Leaving the abundances free to vary produced poorly constrained values.
No clear explanation can be found for this result that
may also be due to the time variability of the absorber.

\section{Evolution of the spectral parameters}

We analyzed the variations in the best-fit spectral parameters
 in the course of the observation to search for possible
correlations between them and with the changes in the mean source luminosity.
We present our results in term of the two main components of the
model: the disk and the corona.

\begin{figure}[h]
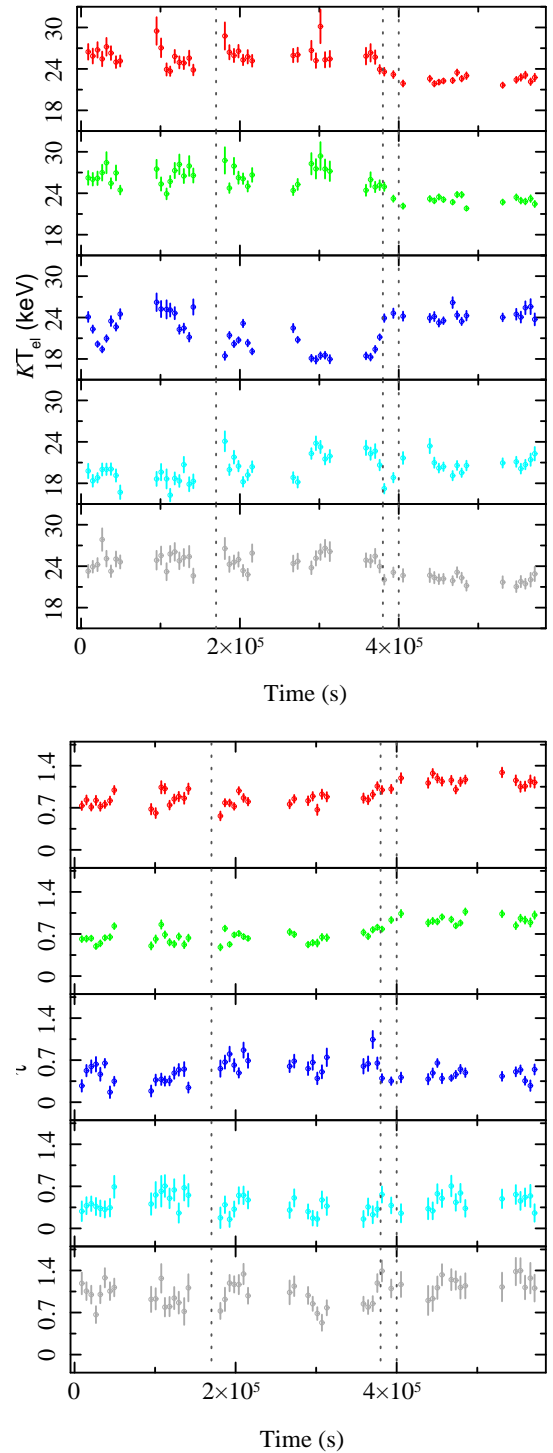

\vspace{0.1 cm}
\centerline{
\vbox{
\psfig{figure=figura6_a.ps,width=7.0cm,angle=0,clip=}
\vspace{0.4 cm}
\psfig{figure=figura6_b.ps,width=7.0cm,angle=0,clip=}
}}
\caption {{\bf Corona}: Temperature (top figure) and optical depth (bottom figure)
of the thermal component of the electron population in the corona vs time
in the five segments.}
\label{corona}
\end{figure}

\begin{figure}[h]
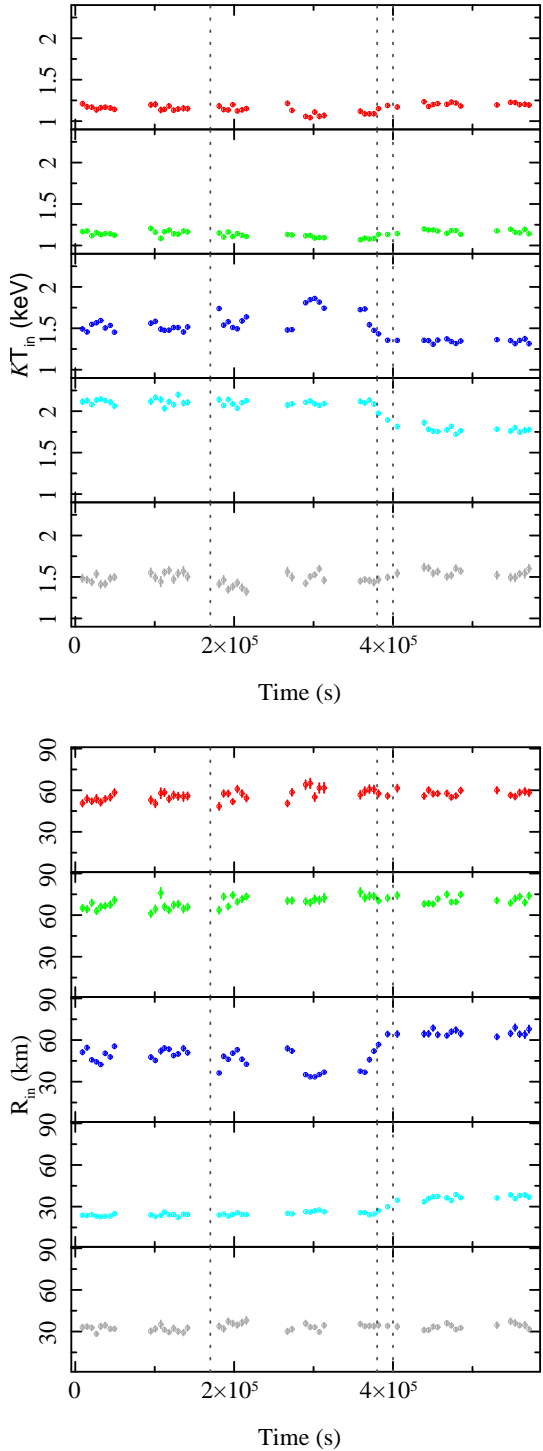

\centerline{
\vbox{
\psfig{figure=figura7_a.ps,width=7.0cm,angle=0,clip=}
\vspace{0.4 cm}
\psfig{figure=figura7_b.ps,width=7.0cm,angle=0,clip=}
}}
\caption {{\bf Multi-temperature black-body disk}: Disk temperature (top
figure) and apparent inner radius (bottom figure) vs time in the five segments.
}
\label{disk}
\end{figure}

For the three intervals we evaluated the average values  of the parameters
of each component and their rms, as given in Table~\ref{table6},  to
take into account the variability, which is also listed in the table as uncertainties.
The time evolution of the parameters for the corona
and the disk component are shown in Figs.~\ref{corona} and
~\ref{disk}, respectively. Moreover, at the end of the fitting
procedure, we derived the disk contribution to the total flux by setting 
$\tau$ and $\Omega$ to zero, and equated the contribution of the corona to the difference.
Non-thermal electrons are responsible for $\sim$30\% in {\it SLTs} 
and 15\% in the {\it pulse} of the coronal emission as evaluated by setting  
the parameter $gmin$ to a negative value (i.e., purely thermal case).

The reflection component was detected with a significance higher
than 3$\sigma$, in the {\it SLT}, while no reflection is observed in the
{\it pulse}. Its contribution to the total flux is always lower than 10\%.
The solid angle has almost no significant variations either with time or segment. 
In particular, we obtained  $\Omega$=1.2$\pm$0.3
and $\Omega$=1.6$\pm$0.3 for {\it SLT-1}  and {\it SLT-2}, respectively,
for the average and rms value during the ten days of observations.
These  values  are higher than  expected based on the
upper limit to the equivalent width of the narrow iron-K line 
($\Omega\leq$0.5), but, when this parameter is constrained to values lower
then the upper limit, the maximum ($\Omega$=0.5)  for all spectra is found.  
Several different values are given in the literature for the reflecting
solid angle, ranging from 0.1 to 2 \citep{ Feroci1999, Zdziarski2005,
Martocchia2006, Ueda2009, vanOers2010}, that could be
related to the spectral state of the source \citep{Sobolewska2003}.
Results obtained from our analysis are compatible with the ones
presented by \citet{Sobolewska2003} and \citet{Zdziarski2005} 
for other variability classes.
However, the low iron abundance that we used for our fitting could lead to
an overestimate of the intensity of the reflection component.

%Figura 6

%Figura 7

The variations in $kT_{el}$, $\tau$, $kT_{in}$, and $R_{in}$
were found to be either  correlated or anti-correlated
with the 1.6-10 keV MECS and 15-100 keV PDS  rates
for the majority of the segments.
These parameters have almost constant values in the intervals
I and III ($regular$ mode) with a significant change in the
mean values occurring in correspondence with the increase
in the count rate. During interval II ($irregular$ mode),
the values are remarkably similar to those of interval I, but
have a larger variance.

\begin{figure}[h]
\centerline{
\vbox{
\psfig{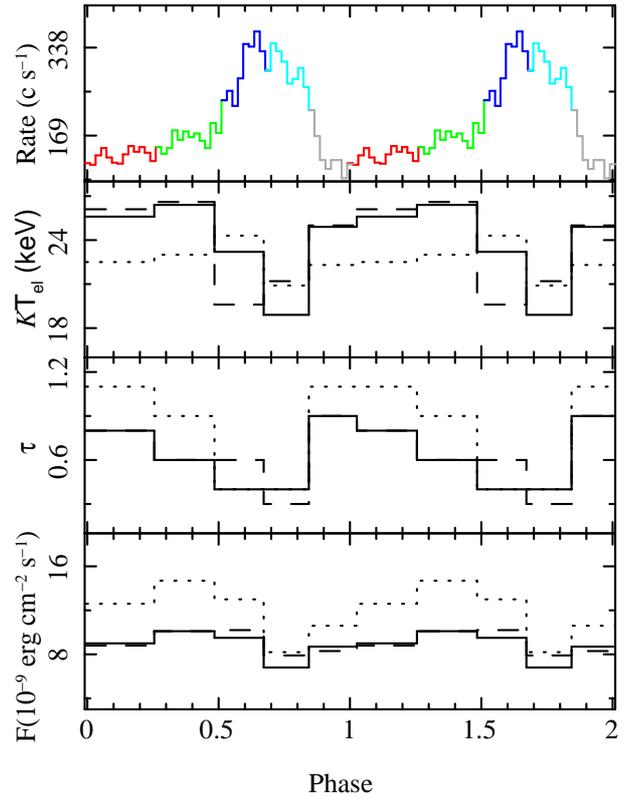}
}}
\caption {{\bf Corona}: Variability of the spectral parameters along
the segments.  
The  electron temperature is shown in the second panel, the optical depth
in the third, and the 0.01-100 keV flux in the bottom one.
Interval I is represented with a solid line, II with a dashed line, and
III with a dotted line.
The top panel shows the MECS rate for comparison; two bursts are shown
for clarity.
}
\label{lc_corona}
\end{figure}

\begin{figure}[h]
\centerline{
\vbox{
\psfig{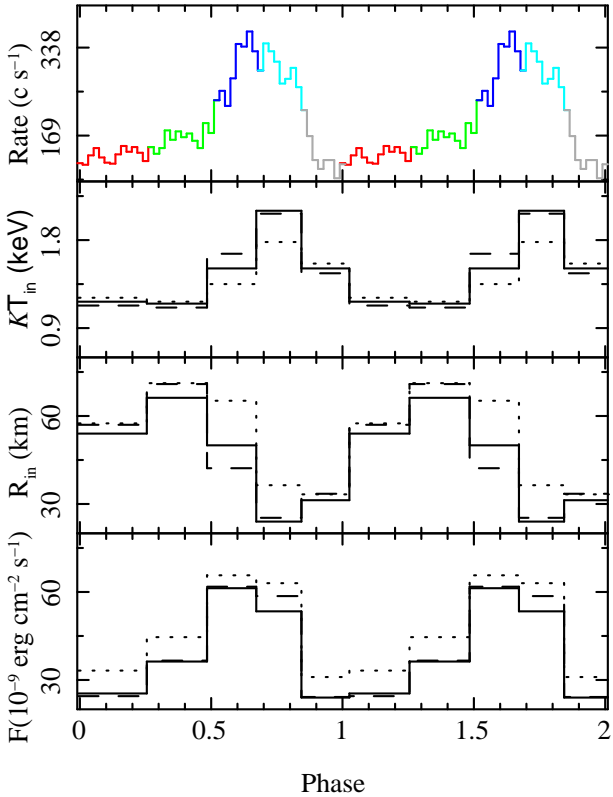}
}}
\caption {{\bf Multi-temperature black-body disk}: Variability of the
spectral parameters along the segments.
The temperature is shown in the second panel, the  apparent inner radius in the third
and the the 0.01-100 keV flux calculated by setting $\tau$ and $\Omega$ to 0 
in the bottom one.
Interval I is represented with a solid line, II with a dashed line and III
with a dotted line. The top panel shows the MECS rate for comparison; two bursts are shown
for clarity. }
\label{lc_disk}
\end{figure}

The strongest variations are detected for the  segments,
as shown in  Figs.~\ref{lc_corona} and ~\ref{lc_disk},
where the averages for the three intervals are plotted with different
line styles (a solid line for interval I, a dashed line for II, and  
a dotted line for III). The figures also show  the 0.01--200 keV flux in units
of 10$^{-9}$ erg cm$^{-2}$ s$^{-1}$  in the bottom panels
computed  by  extrapolating to lower energy the best-fit model.
The parameters have very similar values in {\it SLT-1} and {\it SLT-2},
while they exhibit significant variations in the {\it pulse} with the exception of
$R_{in}$, which varies throughout all segments and reaches a
maximum in {\it  SLT-2} as previously observed by \citet{Vilhu1998}.
The flux of the multi-temperature disk strongly increases
during the {\it pulse}, and simultaneously the corona contribution reduces
significantly  with respect to the values detected in {\it SLT}.
Moreover, the  corona increases its emission by 40-50\% from interval
II to interval III, while smaller variations (20-30\%) are observed in the disk
flux during the three intervals, suggesting that the increase in the rate after  
4.0$\times$10${^5}$ s is caused  mainly by this component.

%Figura 8

%Figura 9

\begin{figure}[h]
\centerline{
\vbox{
\psfig{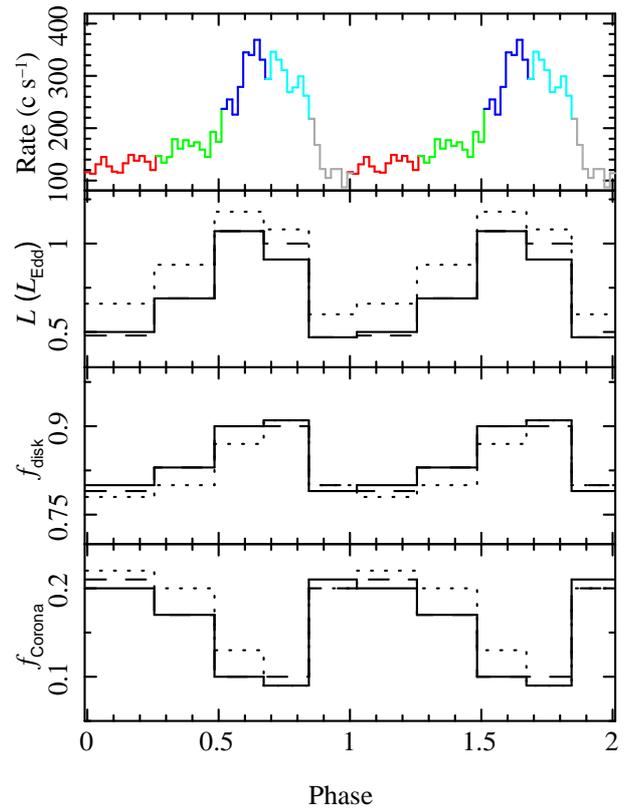}
}}
\caption {Evolution of the total luminosity in units of Eddington
luminosity (second panel) for the considered burst segments.
The fraction of the total luminosity for the disk and
the corona are shown in the third and fourth panels, respectively.
Interval I is represented with a solid line, II with a dashed line, and III
with a dotted line.
The top panel shows the MECS rate for comparison; two bursts are shown
for clarity.}
\label{lumfra}
\end{figure}

The total luminosity of the source, for a distance of 12.5 kpc, was also
computed in the 0.01--200 keV  range as $L_{tot}=L_{disk}+L_{corona}$,
assuming isotropic emission  for both components
and deriving $L_{disk}$  by setting both $\tau$ and $\Omega$ to 0.
The source luminosity changes from $\sim$50\% of the Eddington luminosity
(for a black hole mass of 14 $M_{odot}$)
to 100\% (or slightly above it) in the {\it pulse} as shown
in the second panel of Fig.~\ref{lumfra}. In the third and fourth panels
of the same figure, the fraction of the total luminosity  versus (vs)  segment
for the disk and the corona are also shown.  It is evident that the increase
in the pulse flux is due to the disk contribution,
and that a higher percentage of photons is Comptonized by
the corona after 400 ks.

%Figura 10

% ========= Table 2 =====================

\begin{table*}[h]
\caption{Average best-fit parameters for the {\it Beppo}SAX spectra fit with 
the ``good model'' in the three intervals I (0-170 ks), II (170-380 ks),
and III (400-600 ks). 
The last section of the table shows the results relative to the two RXTE spectra.
}
\label{table6}
\begin{center}
\vbox{
\begin{tabular}{lccc||ccc||c}
\hline\hline
            &\multicolumn{3}{c||}{{\bf Disk}}    &\multicolumn{3}{c||}{{\bf Corona}} &   \\
            & $kT_{in}$ & $R_{in}$ &  F$^*$      &  $kT_{el}$  & $\tau$ &  F$^*$    &  {\bf N$_{\rm H}^{**}$}       \\
            & (keV)    & (km)     &     &  (keV)     &         &  &    \\
\hline\\
\multicolumn{7}{l}{Interval I}    \\
\hline
{\it SLT-1} & 1.17$\pm$0.02 &  54.0$\pm$2.1 & 25.7$\pm$1.3   &  25.6$\pm$1.3 & 0.8$\pm$0.1  &  9.0$\pm$0.9 & 4.64$\pm$0.06  \\
{\it SLT-2} & 1.15$\pm$0.02 &  66.2$\pm$2.5 & 36.8$\pm$1.5   &  26.4$\pm$1.1 & 0.6$\pm$0.1  & 10.1$\pm$1.3 & 4.98$\pm$0.04  \\
{\it P-1}   & 1.51$\pm$0.04 &  50.0$\pm$3.7 & 62.0$\pm$3.3   &  23.2$\pm$1.9 & 0.4$\pm$0.1  &  9.5$\pm$0.7 & 5.27$\pm$0.08  \\
{\it P-2}   & 2.10$\pm$0.07 &  24.0$\pm$1.4 & 53.6$\pm$1.5   &  18.9$\pm$1.1 & 0.4$\pm$0.1  &  6.8$\pm$0.8 & 4.86$\pm$0.08  \\
{\it FDT}   & 1.51$\pm$0.05 &  31.3$\pm$1.8 & 24.0$\pm$1.6   &  24.9$\pm$1.2 & 0.9$\pm$0.2  &  8.7$\pm$0.6 & 4.31$\pm$0.10  \\
\hline \\                                                                                                                                          
\multicolumn{8}{l}{Interval II}                                                                                                   \\
\hline                                                                                                                                            
{\it SLT-1} & 1.13$\pm$0.05 &  57.0$\pm$4.2 & 24.7$\pm$1.7   &  26.1$\pm$1.2 & 0.8$\pm$0.1  &  8.8$\pm$0.8 & 4.6$\pm$0.1    \\
{\it SLT-2} & 1.11$\pm$0.02 &  70.9$\pm$3.1 & 36.7$\pm$1.9   &  26.6$\pm$1.4 & 0.6$\pm$0.1  & 10.1$\pm$0.8 & 5.0$\pm$0.1    \\
{\it P-1}   & 1.66$\pm$0.13 &  42.2$\pm$7.1 & 61.5$\pm$2.8   &  19.6$\pm$1.6 & 0.6$\pm$0.1  & 10.2$\pm$0.7 & 5.2$\pm$0.1    \\
{\it P-2}   & 2.07$\pm$0.07 &  25.3$\pm$1.1 & 58.4$\pm$4.8   &  21.2$\pm$1.8 & 0.3$\pm$0.1  &  7.9$\pm$0.5 & 5.0$\pm$0.1    \\
{\it FDT}   & 1.46$\pm$0.07 &  33.5$\pm$2.3 & 24.2$\pm$2.0   &  25.0$\pm$1.0 & 0.9$\pm$0.2  &  8.3$\pm$0.8 & 4.4$\pm$0.1   \\
\hline\\                                                                                                                                      
\multicolumn{8}{l}{Interval III}                                                                                                                   \\
\hline                                                                                                                                        
{\it SLT-1} & 1.21$\pm$0.02 &  57.5$\pm$2.1 & 33.6$\pm$0.8   &  22.5$\pm$0.5 & 1.1$\pm$0.1  & 12.6$\pm$0.5 & 4.83$\pm$0.07   \\
{\it SLT-2} & 1.17$\pm$0.02 &  71.2$\pm$2.4 & 45.3$\pm$0.9   &  23.0$\pm$0.5 & 0.9$\pm$0.1  & 14.7$\pm$0.8 & 5.10$\pm$0.05  \\
{\it P-1}   & 1.35$\pm$0.02 &  65.2$\pm$2.0 & 67.1$\pm$1.3   &  24.3$\pm$0.8 & 0.4$\pm$0.1  & 13.0$\pm$0.7 & 5.39$\pm$0.05   \\
{\it P-2}   & 1.78$\pm$0.03 &  36.4$\pm$1.5 & 63.3$\pm$2.1   &  20.9$\pm$1.0 & 0.4$\pm$0.1  &  8.2$\pm$0.8 & 5.15$\pm$0.05   \\
{\it FDT}   & 1.56$\pm$0.04 &  33.3$\pm$1.9 & 31.0$\pm$0.8   &  22.3$\pm$0.7 & 1.1$\pm$0.1  & 10.6$\pm$0.6 & 4.53$\pm$0.09   \\
\hline\\                                                                                                                                          
\multicolumn{8}{l}{RXTE}                                                                                                   \\
\hline  
{\it SLT}   & 1.34$\pm$0.02 &  60$\pm$3     & 56.2$\pm$1.0   &  20.5$\pm$0.3 & 1.06$\pm$0.06  & 15.9$\pm$0.3 & 5.7$\pm$0.3    \\
{\it P}     & 1.50$\pm$0.03 &  56$\pm$1     & 74.6$\pm$3.1   &  21.2$\pm$0.4 & 0.60$\pm$0.09  & 15.4$\pm$0.1 & 5.3$\pm$0.3    \\
\hline
\multicolumn{8}{l}{$^*$Flux in the range 0.01-200 keV in unit of 10$^{-9}$  erg cm$^{-2}$ s$^{-1}$} \\
\multicolumn{8}{l}{$^{**}$in unit of 10$^{22}$ cm$^{-2}$ }
\end{tabular}
}
\end{center}
\end{table*}

\section{RXTE simultaneous observation}

\begin{figure}[h]
\vspace{0.1 cm}
\centerline{
\vbox{
\psfig{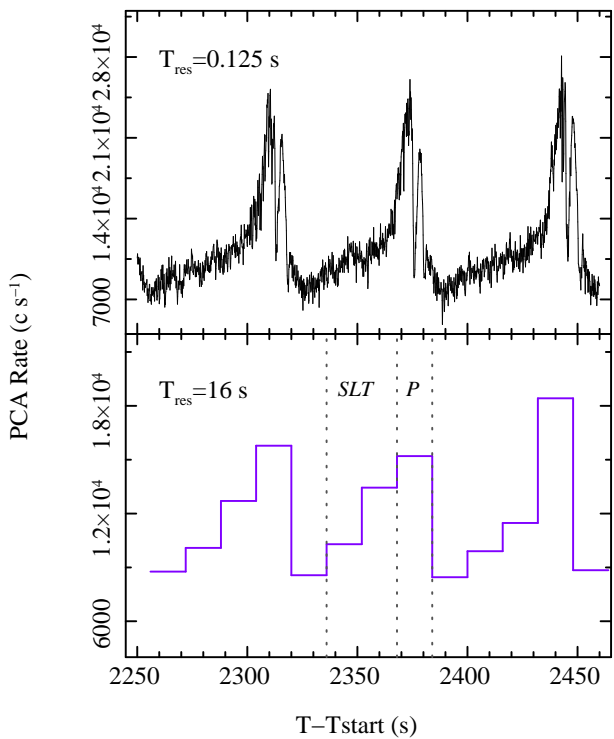}
}}
\caption {Three bursts as observed by the RXTE/PCA  on 2000 October 27
vs time in s from Tstart.
The top panel shows the light curve with 0.125 s time resolution and
the bottom panel the 16 s one.
 Spectral data were only available for the 16 s resolution, and
the vertical dashed lines in the bottom panel indicate the two selection
intervals ($SLT$ and $P$) adopted for the spectral accumulations,
as explained in the text. }
\label{RXTE}
\end{figure}

RXTE observed \grss a few times during the {\it Beppo}SAX long observation.
We extracted the data from observation 50405-01-04-00,
made on 2000 October 27 06:28-09:25 UT.
The starting time of this data series was  550830.0 s after
the beginning of the {\it Beppo}SAX pointing.
The exposure times and count rates were 2480 s and 3033.7 ct s$^{-1}$ for the PCA, and
805 s and 28.3 ct s$^{-1}$ for HEXTE, respectively.
Spectral data were available only with time resolution of 16 seconds,
we then defined from the PCA light curve (see bottom panel of Fig.~\ref{RXTE})
two intervals, {\it SLT} (9000-15000 ct s$^{-1}$ ) and {\it  P} ($>$15000  ct s$^{-1}$),
where were as close as possible to the {\it Beppo}SAX finer selection.
Two spectra were extracted from both the PCA (PCU2 only) and HEXTE instruments,
following standard RXTE procedures and a 1.6\% systematic error was added to the PCA data.
Table~\ref{table6} shows the best-fit parameters relative to the ``good model''
that are in good agreement with the {\it Beppo}SAX ones.
We note that the high-resolution light curve, plotted in the top panel of
Fig.~\ref{RXTE}  shows a deep ``dent'' in the bursts \citep[see][]{Belloni2000}.
This is true for all RXTE bursts observed during the {\it Beppo}SAX interval, but
since no spectral information was available for these data,
it was impossible to perform a more time-resolved analysis.

%Figure11

\section{Discussion}

The phenomenological model used to fit all spectra, has allowed us to derive some
physical information about the characteristics of the source emission in a $\rho$ mode.
The scenario that we adopt in the following discussion is that of an accretion disk plus
a surrounding high temperature corona.
These two components and their strong interaction with each other have been proposed
by several authors \citep{Taam1997,Zdziarski2001,Done2004,Ueda2009} in
describing the complex behaviour of \grs.

\subsection{The multi-temperature disk parameters}

The apparent inner radius and temperature are the two most important 
parameters of the disk component when attempting to interpret   
the emission of  \grss.
The values of $kT_{in}$ found in our analysis vary within the range
1 -- 2 keV and those of  $R_{in}$ are in the range 20--70 km.
These results compare quite well with other results for  $\rho$ \citep{Vilhu1998}
and other classes \citep{Belloni1997}, although temperatures as low as 0.7 keV
have also been reported \citep{Fender2004}.

According to the Shakura-Sunyaev thin disk model, $R_{in}$ depends only on
the mass and spin of the black hole.  
Our results are compatible with the gravitational radius ($R_g = GM / c^2$) of a
14 $M_{\odot}$ black hole, which is equal to $\sim$21 km.
As pointed out by \citet{Vierdayanti2010}, to evaluate the $true$ inner radius  
$R_{in}^*$ we must consider the relativistic and spectral hardening corrections,
which can be factorized as \citep{Kubota1998}
\begin{equation}
R_{in}^* = \kappa^2 \zeta R_{in}
\end{equation}
\noindent
adopting their values ($\kappa=1.7$, $\zeta=0.412$), we obtain values larger
than $R_g$ within the reported uncertainties.

In any case, our lower values of $R_{in}^*$ are smaller than the Schwarzschild
radius $R_S = 2 R_g$. This might be an indication of either a rapidly rotating black
hole or  a slim accretion disk at about the Eddington luminosity.
Indications of a spinning black hole in \grss have been presented by several
authors  \citep[][and references therein]{Middleton2006},  who estimated
a dimensionless spin parameter $a = Jc/GM^2$, where $J$ is the angular
momentum of the black hole, equal to about $\sim$0.7.
With this value, a minimum distance of $\sim$36 km is acceptable, and considering
the large uncertainty in the mass, which might be slightly overestimated, our results
are not in conflict with the general scenario adopted for this peculiar source.
On the other hand, the calculations of \citet{Watarai2000} for the emission from
a slim disk, demonstrated that radiation can be produced at distances much smaller than
the last stable orbit, i.e. in the region between $R_S$ and $3 R_S$.

\begin{figure}[h]
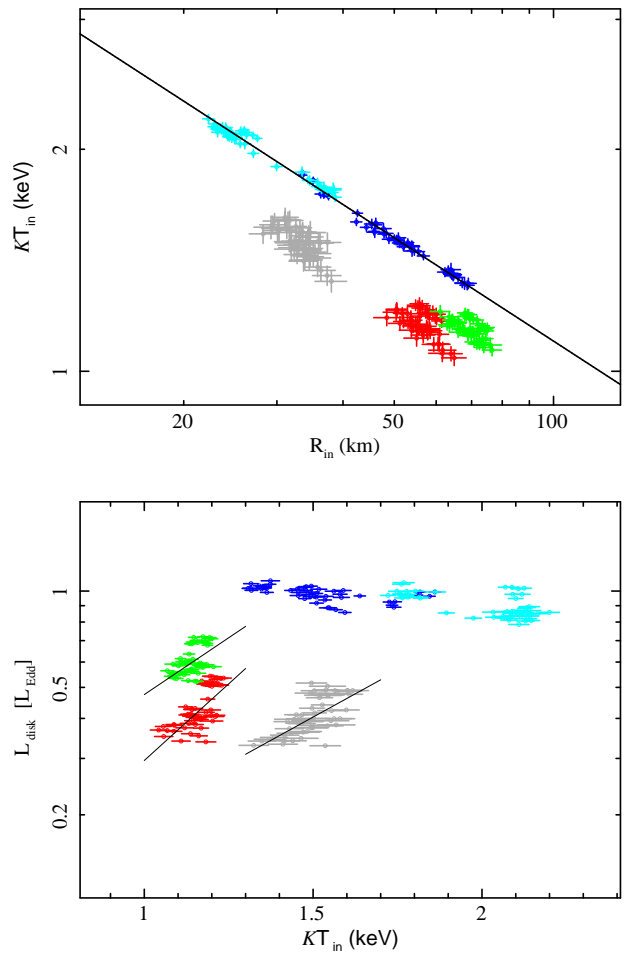

\vspace{0.2 cm}
\centerline{
\vbox{
\psfig{figure=figura12_a.ps,width=8.0cm,angle=-90,clip=}
\vspace{0.5cm}
\psfig{figure=figura12_b.ps,width=8.0cm,angle=-90,clip=}
}}  
\caption {{\it Upper panel}: temperature of the multi-color disk vs apparent inner radius.
{\it Lower panel}: disk luminosity, in units of Eddington luminosity, vs
the inner temperature. The continuous black lines are the best-fit models (see text).
}
\label{tin_rin}
\end{figure}

The plot of $T_{in}$ vs $R_{in}$ for the multi-temperature disk  obtained
for all the fitted spectra is shown in the top panel of Fig.~\ref{tin_rin}.
The points of {\it SLT-1} and {\it SLT-2} form two clusters with the same
temperature range but clearly separated in terms of $R_{in}$, because of the different
flux levels: both clusters exhibit an anticorrelation between the two
parameters.
The tight correlation is clearly evident between {\it P-1}  and  {\it P-2}, and
appears to behave coherently during the entire observation.
We fitted the parameter values for these two sets with a single power-law
\begin{equation}
T_{in} = C R^{-q}_{in},
\end{equation}
\noindent
leaving the exponent $q$ as a free parameter and obtained $q=$0.46$\pm$0.01;
the fitting relation is shown in Fig.~\ref{tin_rin} with a solid line.
This result translates into an almost constant flux at the peak,
as shown in  the bottom panel of Fig.~\ref{tin_rin}, where the disk luminosity for
all the spectra is plotted vs $kT_{in}$.

%figura12

Observations of several BHCs have shown that whenever
changes have been detected, the luminosity of the disk varies in proportion
to $T^4$ \citep[][and reference within]{Done2007}.
However, deviations from this relation of the form $L \propto T^2$ have been observed in
some objects, as in the case of XTE J1550-564, which has an inclination angle of
$\sim$70$^{\circ}$ and a mass 8--11 $M_{\odot}$, and exhibits superluminal jets
\citep{Kubota2004a,Kubota2004b}, similar to \grs.
Fitting points from all segments, except for that containing the {\it pulse}, with a power-law,
the resulting values for the temperature exponent are
 2.7$\pm$0.4 for {\it SLT-1}, 2.0$\pm$0.5 for {\it SLT-2}, and 1.9$\pm$0.2 for
{\it FDT}, respectively.
Several models have been proposed to explain the relation $L \propto T^2$ for
a slim disk where advection dominates the energy transfer in the inner region and
reduces the radiation efficiency \citep{Watarai2000}.
Moreover, the effects of the large inclination angle cannot be neglected since
deviations from the $L \propto T^4$ relation are mainly observed in highly inclined
sources \citep{Done2007}.

\subsection{The corona}
\begin{figure}[h]
\centerline{
\vbox{
\psfig{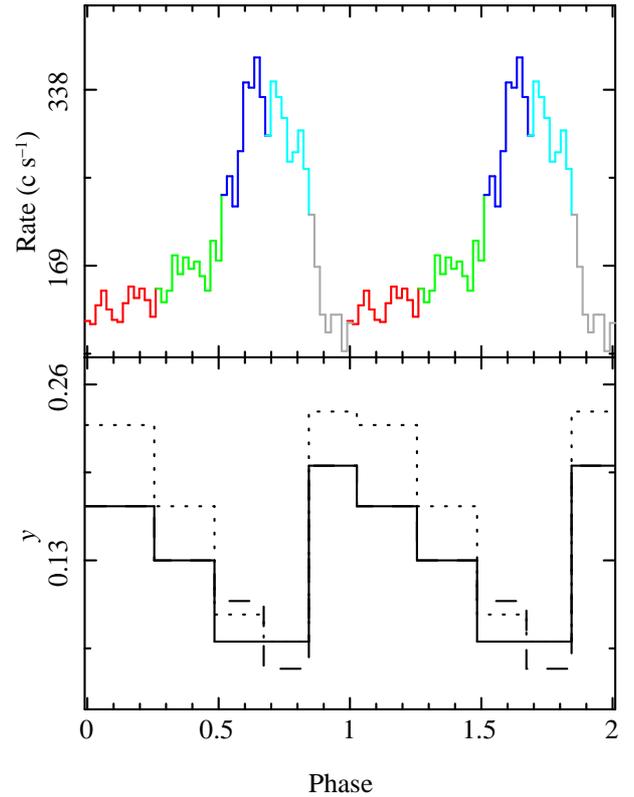}
}}
\caption {{\bf Corona}: Variability of the  Comptonization parameters $y$ with
the burst segments.  
Interval I is represented with a solid line, II with a dashed line, and
III with a dotted line.
The top panel shows the MECS rate for comparison; two bursts are shown
for clarity.
}
\label{compatt}
\end{figure}

The emission of the corona has been detected up to 600 keV without any
break  and can be interpreted in terms of both a thermal and a non-thermal 
electron population \citep{Zdziarski2001}.
The output spectrum of the thermal component, which is mainly responsible
for the spectrum in the {\it Beppo}SAX band, depends on the Comptonization
parameter $y$, defined as \citep{Rybicki1979}

\begin{equation}
y = 4 \frac{kT_{el}} {m_{e} c^{2}} {\rm Max} (\tau,\tau^2), 
\end{equation}

\noindent
where $m_e$ is the electron mass and $c$ the speed of light.
The values of $y$ vs segment are plotted in Fig.~\ref{compatt};
no large variations are observed  between the time intervals
I and II (continuous and dashed lines), while interval III is characterized by higher
values, in agreement with the results for the absolute flux
emitted by the corona.
Significant variations in $y$ are observed for the segments with higher values in
both $SLT$ and $FDT$, while these variations are smaller  than  0.1 in the 
$pulse$, in accordance with the trend in the flux.

%figura13

As shown by \citet{Sunyaev1980}, the ratio of the luminosity of the Comptonizing
corona $L_c$ to the luminosity of the source of the seed photons ($L_0$) for
$kT_e>h\nu$ is a function of the spectral index, thus for two different segments
we have
\begin{equation}
\frac{L_{cj}}{L_{ci}}=\frac{L_{0j}}{L_{0i}}\times f(\Gamma_{j},\Gamma_{i}),
\end{equation}
\noindent
where $i$ and $j$ indicate different  segments and $\Gamma$ is
the energy index.
The spectral index values obtained for the model $v$,
are practically constant with $\Gamma\simeq$1.6 (see model $v$ in Appendix A),
hence the function  $f(\Gamma_{j},\Gamma_{i})$ is close to unity.
The decrease in the corona luminosity observed at the peak section
could then be  attributed to a reduction in the size of the corona itself.
Models of disk+corona systems have shown that  
strong interactions between the two components are able to explain the
constant spectral index in term of a tight interaction between
the temperature and the optical depth of the corona
\citep{Haardt1991}. Moreover, the thermal coupling of the disk and the corona
would result in either the evaporation of the disk or the condensation of the corona
depending on the accretion rate \citep{Liu2011,Meyer2007,Haardt1993}.

\section{Summary and conclusion}

The spectra of GRS1915+105 that we have accumulated in five segments  of the burst of the
$\rho$ class have been modeled with several combinations of spectral laws.
Only two models were found to provide acceptable fits to all spectra: a multi-color disk 
with both a black-body and power law, and a disk plus an hybrid corona.
Since the former does not have a consistent physical interpretation,
we studied the spectral changes in the source based on the basis of the latter.

The following conclusions can be inferred from the analysis
of the best-fit values of the parameters:

\begin{itemize}
\item
The total 0.01-200 keV luminosity  of the
source in all segments, although the {\it pulse} displays a small variation
with time in the first 300 ks and sharply increases to a second almost constant
level in the last 200 ks.
This step-like behaviour is found for most of the spectral parameters,
with opposite trends.
The luminosity of the {\it pulse} remains remarkably stable with time,
in the MECS range (1.6-10 keV) despite the variation in the relative
best-fit parameters.
\item
The disk luminosity increases in the {\it pulse} and the
$R_{in}-T_{in}$ correlation can be more accurately interpreted 
using the slim disk model.
\item
The  luminosity of the corona is lower at the bursts,
possibly because the corona condenses onto the disk.
\end{itemize}

We stress that some of these results are model-independent
being mainly related to the observed rate. All models
that we have investigated show that the luminosity below 10 keV
reaches the same level in all runs at the peak, whereas at the
same time the flux  above 20 keV decreases respect to
the $SLT$. This behaviour should  be taken into
account  in any more detailed modelings of the complex emission
of this source.

\begin{acknowledgements}

The authors thank the anonymous referee for improving the 
paper with constructive comments and suggestions. 
They thank the personnel of ASI Science Data
Center, particularly M. Capalbi, for help in retrieving \sax
archive data. TM thanks A.A. Zdziarski for 
useful discussions about the analysis.
\\
This work has been partially supported by research
funds of the Sapienza Universit\`a di Roma. The INAF Institutes
are financially supported by the Italian Space Agency (ASI) in the
framework of the contracts ASI-INAF I/023/05/0 (M.F.) and ASI-INAF
I/088/06/0. PC acknowledges support from a EU Marie Curie
Intra-European Fellowship under contract no. 2009-237722

\end{acknowledgements}

\bibliographystyle{aa}

\bibliography{grs1915}

\appendix
\section*{Appendix A: Spectral models}
We summarize the results obtained with several combinations of
spectral functions in XSPEC, which did not provide acceptable results for all segments.

\indent
$i$) -- {\it Multi-temperature black-body disk + power law}:
this is one of the simplest models frequently adopted in the spectral analysis of
\grss observations \citep{Belloni1997, Taam1997}.
We obtained acceptable reduced $\chi^2$ only for a few spectra and, in all data series,
large residuals were found to be apparent above 10 keV, which cannot be due to instrumental errors
because they were not eliminated by a change in the intercalibration factors
within the acceptable range.
Adding an exponential cut-off to the power law improves the fits giving narrower
reduced $\chi^2$ distributions, which are marginally compatible ($\sim$2.5 Gaussian
$\sigma$ far) with the expected one for the spectra in the {\it SLT-1}, {\it SLT-2}, and
{\it FDT} segments  where an average spectral index of 2.4 and a cut-off at 60 keV are
measured.
However, large residuals above 30 keV in the {\it P-1} and {\it P-2} spectra, caused by 
the underestimation of the flux, produce  distributions of the reduced $\chi^2$ of
negligible probability of being equivalent to the expected one.

$ii$) -- {\it Multi-temperature black-body disk + Comptonised spectrum}:
the change in the cut-off power law with a comptonization spectrum \citep{Sunyaev1980}
from high temperature electrons ({\sc compst}) provides the same results as the previous
model.

$iii$) --{\it Multi-temperature black-body disk + power law + reflection from accretion
disk}:
according to previous results \citep{Feroci1999,Martocchia2002}, we added the Compton
reflection from an accretion disk that is either cold ({\sc pexrav}) or ionized ({\sc pexriv})
\citep{Magdziarz1995} to the multi-temperature black-body disk plus power-law model.
These models did not originally include any iron emission line which was added with a Gaussian
of fixed width (0.1 keV).
Acceptable fits were obtained for the {\it FDT} interval, where the spectral
statistics are the lowest, and in the {\it SLT-1} spectra before the flux
increase at 380 ks; the $\chi^2$ distribution in these cases is compatible to within
2$\sigma$ with the expected one.
The two models {\sc pexriv} and {\sc pexrav} provide compatible best-fit parameters, and in
particular the ionization parameter in {\sc pexriv} assumes values compatible with zero.
In the other segments, the model is  unable to  describe the data above 15 keV, where
the highest residuals are observed.

$iv$) -- {\it Comptonization from bulk-motion matter}:
the model ({\sc BMC} in XSPEC) was developed by \citet{Titarchuk1997} 
to describe the emission of accreting matter onto a
black hole and takes into account the Comptonization of soft photons by a hot gas 
and the gain in energy originating from the dynamic bulk motion.
The combination of two {\sc BMC} components, which was used  by 
\citet{Titarchuk2009} to fit many RXTE spectra
of \grs, is characterized by the black-body temperature of the soft photon source, a
spectral (energy) index, and an illumination parameter related to the fraction of the
bulk-motion flow irradiated by the thermal photon source.
We left the temperatures of the two {\sc BMC} as free parameters,
because fixing them to the value used by \citet{Titarchuk2009} for the RXTE data
(1 keV) did not provide acceptable fits to all spectra.
An iron line with a Gaussian shape was added to the model, together with an high energy
cut-off for one of the two {\sc BMC} components.
Once again, we found that the best-fit spectral distributions, particularly those of the
{\it P-1} and {\it P-2} intervals, were not acceptable, in term of either the $\chi^2$ values
or the large residuals.

The use of an equivalent column density, $N_H$ = 5$\times$10$^{22}$ cm$^{-2}$, as the one
already considered by \citet{Titarchuk2009}, provided only a marginal improvement
to the fits.
The spectra from the {\it SLT-1} and {\it FDT} intervals could be modeled by a combination
of two {\sc BMCs},  the $\chi^2$ distribution of the spectra were compatible at the 2$\sigma$ 
level with the expected ones, and the spectral parameters of the first {\sc BMC} are 
well-defined with an approximately stable temperature ($kT \simeq$ 1 keV), photon index 
($\Gamma$ = 2.8), and Comptonised fraction ($f$ = 0.5).
The second {\sc BMC} has a flux about 60\% fainter than the first and a lower temperature
($kT \simeq$ 0.5 keV), but the photon index and the Comptonised fraction are not confined
for almost all spectra.
Spectra for the {\it SLT-2} intervals are marginally compatible with the model,
 their distribution $\sim$2.5 Gaussian $\sigma$ being far from that expected,
while large residuals above 20--30 keV are present in the pulse spectra
(see Fig. 5).
Moreover, the $\chi^2$ distribution is not compatible with the expected one.

$v$) --{\it Multi-temperature black-body disk + power law + black body}:
To reduce the large deviations in the residuals between 10 and 30 keV
obtained with the other composite phenomenological models,
we added to the multi-temperature black-body disk plus power law
a third spectral component whose emission is mainly concentrated in this
limited band.
This model cannot be rejected based on statistical arguments
giving acceptable reduced $\chi^2$ and flat residuals for all spectra in
all segments.
The results for the disk  are  in close agreement
with those relative to the ``good model''. The power law
shows a flux that decreases at the peak coherently with the
results of the disk+hybrid corona model and the
spectral index displays small variations ($<$10\%) with both time and segment,
having  a mean value of $\alpha$=2.6.
The black-body temperature ranges between 3 and 6 keV with the
highest values being observed during the $SLT$;
its contribution  to the source luminosity is always very low, but  becomes significant
at energies higher than $\sim$ 10 kev, particularly in the {\it pulse}, when the
power law is faint.
No correlation with the power-law luminosity is observed during the bursts, and
we note moreover that during these periods it subtends a different spectral range,
across which the temperature decreases to 3 keV.
However, a consistent scenario to explain the nature and the changes in this component
cannot be easily envisaged.
\citet{Titarchuk2009} also found a similar component with a
temperature of 4--5 keV in some of the spectra observed with RXTE and proposed an
interpretation in terms of a highly gravitationally redshifted annihilation line.
Such a high redshift can be reached only if the emission region is at an  extremely
short distance from the Schwarzschild radius, and therefore it
seems quite unlikely that the flux would exhibit the same modulation of the
disk.

\end{document}